\pdfoutput=1
\documentclass[prd,nofootinbib,showpacs,superscriptaddress,preprint]{revtex4}
\usepackage[T1]{fontenc}
\usepackage{amsmath,amssymb}
\usepackage{epsfig}
\usepackage{dcolumn}
\usepackage{graphicx}
\usepackage[usenames,dvipsnames]{color}
\usepackage{slashed}
\usepackage[colorlinks,citecolor=blue]{hyperref}
\usepackage{pdfpages}
\usepackage{float}
\usepackage{soul}
\begin{document}
	\title{Impact of texture zeros on dark matter and neutrinoless double beta decay in inverse seesaw}
		
	\author{Nayana Gautam}
	\email{nayana@tezu.ernet.in}
	\affiliation{Department of Physics, Tezpur University, Tezpur - 784028, India}
	
	\author{Mrinal Kumar Das}
	\email{mkdas@tezu.ernet.in}
	\affiliation{Department of Physics, Tezpur University, Tezpur - 784028, India}
	
\begin{abstract}
	We study the impact of maximal zeros of the Dirac mass matrix on neutrino phenomenology and dark matter within the framework of an inverse seesaw ISS $(2,3)$. ISS $(2,3)$ is obtained by adding two right handed neutrinos and three gauge singlets sterile fermions to the standard model leading to an extra sterile state in the keV range. The model is more predictive because of the presence of fewer numbers of right handed neutrinos than the conventional inverse seesaw. Moreover, texture zeros in the structures of the mass matrices involved in the model can reduce the free parameters. We extensively study the effect of different textures of Dirac mass matrix on sterile neutrino dark matter phenomenology. Our study also includes the implications of these texture zero mass matrices on low energy phenomena like neutrinoless double beta decay (0$\nu\beta\beta$). The zero textures highly constrain the parameter space of the model. Based on the allowed cosmological ranges of the relic abundance, decay rates and dark matter mixing with the active neutrinos, we study the viability of the different textures within the framework. 

	\end{abstract}
\maketitle

\section{\label{sec:level1}Introduction}
In spite of the robust progress in our knowledge of neutrino masses and lepton mixing \cite{Mohapatra:2005wg,Ma:2016mwh,King:2008vg,Mohapatra:1986bd,Schechter:1980gr,Mohapatra:1980yp,Wetterich:1981bx}, there are still some unsolved issues in the light neutrino sector. Though there are precise measurements of the mixing angles and mass squared differences (one can see table \ref{tab3}), yet there are no conclusive remarks on the Dirac CP phase ($\delta_{CP}$). Again, large mixing of neutrinos, ordering of the three neutrino masses, uncertainties in the octant of atmospheric mixing angle are major issues in the physics of light neutrinos. In addition to these, dark matter (DM) has been a great mystery in particle physics and cosmology to date \cite{Zwicky:1933gu,Clowe:2006eq,Battaglieri:2017aum}. Since the standard model (SM) cannot provide a viable dark matter candidate, there is a need for extension of the standard model by the addition of other particles which can have a possibility to account for dark matter. The inclusion of gauge singlet sterile fermions to the standard model has been highly motivated to explain dark matter along with the active neutrino phenomenology \cite{Boyarsky:2018tvu,Abada:2017ieq}. As proposed in \cite{Adhikari:2016bei}, sterile neutrinos in keV scale can be a well motivated feasible dark matter candidate. Sterile neutrino is a singlet right handed neutrino having tiny mixing with the SM neutrinos and have a very long lifetime \cite{Boyarsky:2008mt}. However, there are some cosmological bounds to be satisfied by sterile neutrino dark matter which can be found in \cite{Taoso:2007qk,Ng:2019gch,Abazajian:2012ys}. 

There are several beyond standard model (BSM) frameworks proposed to explain these issues of the standard model. Inverse seesaw is one of the most followed frameworks that can address the origin of neutrino masses at a very low scale compared to the conventional seesaw \cite{MA1987287,Wyler:1982dd,Mohapatra:1986bd,Deppisch:2004fa,Dev:2009aw}. Besides, it can explain dark matter and baryon asymmetry of the universe \cite{Fukugita:1986hr}.  In this work, we have selected an inverse seesaw ISS(2,3) proposed by \cite{Abada:2014vea} where two right handed neutrinos and three sterile fermions are added as our previous work \cite{Gautam:2019pce}. As mentioned in \cite{Abada:2014zra,Gautam:2019pce}, one can naturally obtain a sterile state in keV scale which accounts for the dark matter within the framework of ISS(2,3). Apart from this, another interesting motivation for this model is its viability as it can have significant collider signatures because of this low energy scale. Other phenomenological consequences of this model can be probed at experiments in near future.

It is worth mentioning that ISS (2,3) with the implementation of an additional discrete symmetry constrain the model, enhancing its predictability and testability perspectives, especially concerning its flavor structure and CP properties \cite{Gautam:2019pce}. It is quite interesting that the model augmented with an additional flavor symmetry can account for neutrino as well as DM phenomenology. However, if the mass matrices in the framework have very specific structures with some zero entries at certain positions, then the number of free parameters in the model can be significantly reduced \cite{Ghosal:2015lwa,Kageyama:2002zw,Branco:2007nb,Choubey:2008tb,Borgohain:2020now}. In such a case, we can have very specific predictions for light neutrino parameters \cite{Ghosal:2015lwa}. The idea of implementing texture zeros in inverse seesaw has already been explored by several groups \cite{Ghosal:2015lwa,Adhikary:2013mfa,Sinha:2015ooa}. In these papers, the authors have proposed phenomenologically viable zero textures of the mass matrices within the inverse seesaw framework. However, the advantage of minimal inverse seesaw (ISS(2,3)) over the conventional inverse seesaw is that apart from neutrino phenomenology, one can study and constrain the texture zero of mass matrices from dark matter phenomenology as well which is the main aim of this paper. In ISS(2,3), three mass matrices namely Dirac neutrino mass matrix $M_{D}$, heavy neutrino mass matrix $M_{N}$, and sterile neutrino mass matrix  $\mu$ play role in generating light neutrino mass matrix. It is to note that in this model, $M_{D}$ is a non-squared matrix because of the presence of an unequal number of left handed and right handed neutrinos. It has already been shown in the literature that in the diagonal charged lepton basis, the maximum number of zeros allowed in the light neutrino mass matrix is two. However, among the fifteen possible two zero textures of light neutrino mass matrix, only six are put up with the neutrinos and cosmology data \cite{Meloni:2014yea,Alcaide:2018vni,Bora:2016ygl,Fritzsch:2011qv}. These conditions in the light neutrino mass matrix can constrain the texture zeros in these three matrices $M_{D}$,$M_{N}$, and $\mu$. In our study, we have taken maximal possible zeros in $M_{N}$ as well as in $\mu$, and the vanishing elements are postulated in the matrix $M_{D}$. In this work, we are not addressing the explicit origin of such textures of $M_{D}$, assuming that some underlying flavor symmetry can lead to the given textures \cite{Grimus:2004hf,Borgohain:2018lro}. After listing the possible zero textures of $M_{D}$, we perform a detailed analysis of sterile neutrino dark matter in all the textures. We calculate mass and DM-active mixing, relic abundance and the decay rate of the lightest sterile neutrino in the textures. We study the predictions of all the textures on these parameters related to dark matter. The viability of the different textures is identified by implementing astrophysical and cosmological bounds on dark matter. Apart from dark matter phenomenology, we explore the new physics contribution of the textures to neutrinoless double beta decay (0$\nu\beta\beta$) \cite{Faessler:2014kka,Blennow:2010th}.   

The paper is structured as follows. In section \ref{sec:level2}, we describe the framework of ISS $(2,3)$. In section \ref{sec:level3}, we classify the different possible texture zeros in the framework of ISS $(2,3)$. Section \ref{sec:level4} explains the dark matter production in ISS $(2,3)$. In section \ref{sec:level5}, we briefly discuss about neutrinoless double beta decay process in presence of heavy sterile neutrinos. Section \ref{sec:level6} contains the results of the numerical analysis and discussion. We discuss some laboratory searches for sterile neutrinos in section \ref{sec:level7}. Finally, in section \ref{sec:level8}, we present our conclusion.
\begin{table}[H]
	\centering
	\begin{tabular}{|c|c|c|}
		
		\hline 
		Oscillation parameters	& 3$\sigma$(NO) & 3$\sigma$(IO) \\ 
		\hline 
		$\frac{\Delta m_{21}^{2}}{10^{-5}eV^{2}}$	& 6.82 - 8.04 & 6.82 - 8.04  \\ 
		\hline 
		$\frac{\Delta m_{31}^{2}}{10^{-3}eV^{2}}$	& 2.431 - 2.598  & 2.31-2.51 \\ 
		\hline 
		$sin^{2}\theta_{12}$ &0.269 - 0.343  & 0.269 - 0.343 \\ 
		\hline 
		$sin^{2}\theta_{23}$ &  0.407 - 0.618  &  0.411 - 0.621 \\ 
		\hline 
		$sin^{2}\theta_{13}$ &  0.0203 - 0.0243 &  0.0205 - 0.0244 \\ 
		\hline 
		$\frac{\delta}{\pi}$ & 0.87 - 1.94 &  1.12- 1.94\\ 
		\hline 
	\end{tabular} 
	\caption{Latest Global fit neutrino oscillation Data \cite{deSalas:2020pgw}.}\label{tab3}
\end{table}

\section{\label{sec:level2}Inverse Seesaw framework : ISS(2,3)}
As discussed in several earlier works, inverse seesaw(ISS) is a very well motivated and widely studied BSM framework. This framework ensures the generation of neutrino mass at a very low scale compared to the conventional seesaw which may be probed at LHC and in future neutrino experiments. Three standard model gauge singlet fields are introduced
here along with their Majorana mass matrix having ultralight eigenvalues in the keV scale \cite{Abada:2014zra,Samanta:2015oqa}. The relevant Lagrangian for the ISS is given as,

\begin{equation}
L = - \dfrac{1}{2}{n_{L}^{T}C M n_{L}} + h.c
\end{equation}
Here, charge conjugation matrix C can be written as $C \equiv i\gamma^{2}\gamma^{0}$ and the basis is $n_{L}=(\nu_{L,\alpha},\nu_{R,i}^{c},s_{j})^{T}$. $\nu_{L,\alpha}$ with ($\alpha= e$,$\mu$,$\tau$),$\nu_{R,i}^{c}$ and $s_{j}$ are the SM left handed neutrinos, right handed (RH) neutrinos and sterile fermions respectively. This Lagrangian gives rise to the following neutrino mass matrix:
\begin{equation}\label{eq:2}
M =\left(\begin{array}{ccc}
0 & M_{D}& 0\\
{M_{D}}^{T} & 0 & M_{N} \\
0 & M_{N}^{T} & \mu 
\end{array}\right) 
\end{equation}
where $M_{D}$, $M_{N}$ and $\mu$ are complex matrices. The effective neutrino mass matrix for the active neutrinos can be obtained by the block diagonalisation of the above mass matrix as:
\begin{equation}\label{eq:1a}
M_{\nu}\approx M_{D}^{T}(M_{N}^{T})^{-1}\mu M_{N}^{-1} M_{D}
\end{equation}
One can obtain sub-eV scale the light active neutrinos considering $M_{D}$ at electroweak scale, $M_{N}$ at TeV scale and $\mu$ at keV scale as explained in many literatures \cite{Dias:2011sq,Ma:2009gu}. The remaining heavy part in this model can be written as, 
\[
M_{H} = \begin{pmatrix}
0 & M_{N} \\
M_{N}^{T} & \mu \\
\end{pmatrix}
\] 
Since in ISS $(2,3)$, only two RH neutrinos are involved, M is a $8\times8$ matrix which can further be diagonalised using  $\mathcal{U}$ as,
\begin{equation}\label{eq:1} 
\mathcal{U}^{T}M\mathcal{U} = M^{diag} = diag(m_{1},m_{2}....m_{8})
\end{equation}
Here, $M_{N}$ is a non squared $2\times3$ matrix. So, $M_{N}^{-1}$ in Eq. \ref{eq:1a} is not well defined in this model. Hence, the light neutrino mass matrix can be obtained using the following form proposed by the authors \cite{Abada:2014zra,Abada:2014vea}.
\begin{equation}\label{eq:1b}
M_{\nu}\approx M_{D}d M_{D}^{T} 
\end{equation} 
where d is $2\times2$ dimensional submatrix defined as
\[
{M_{H}}^{-1} = \begin{pmatrix}
d_{2\times2} & .... \\
..... &.... \\
\end{pmatrix}
\]
with 
\[
M_{H} = \begin{pmatrix}
0 & M_{N}\\
M_{N}^{T} & \mu \\
\end{pmatrix}
\]
Active-sterile mixing or DM-active mixing is a crucial ingredient to study dark matter phenomenology within ISS $(2,3)$ \cite{Abada:2014zra}. We have calculated these mixing angles by numerically diagonalising M using Eq.\ref{eq:1}.

\section{\label{sec:level3}Texture Zeros in Dirac Mass Matrices Of ISS(2,3)}

ISS(2,3) model has the advantage in that it depends on fewer parameters than the conventional inverse seesaw model because of the presence of fewer number of right handed neutrinos. Again, implementing texture zeros to the mass matrices involved in the model can reduce the number of model parameters thereby increasing the predictive nature of the model. In this model, three mass matrices $M_{D}$, $M_{N}$ and $\mu$ lead to a light neutrino mass matrix. The allowed number of zeros in the light neutrino mass matrix can constrain the texture zeros in these three matrices. In our study, we have taken maximal possible zeros in $M_{N}$ and $\mu$. $M_{N}$ contains two quasi-Dirac particles with mass hierarchy leading to the fact that there must be at least two non zero entries in $M_{N}$. Again, we have added three sterile fermions in the model and the minimum number of non zero elements in $\mu$ is three. Considering these two crucial points and using discrete flavor symmetry, one can obtain the maximal zeros in $M_{N}$ and $\mu$ as \cite{gautam2021neutrino,Camara:2020efq} ,
\begin{equation} \label{eq:u1}
M_{N}= \left(\begin{array}{ccc}
	f & 0 & 0\\
	0 & g & 0 \\ 
\end{array}\right), \;  \mu = \left(\begin{array}{ccc}
	p & 0 & 0 \\
	0 & p & 0 \\ 
	0 & 0 & p
\end{array}\right)
\end{equation}

Here, we have considered identical parameters in $\mu$ to keep the parameters in the model as minimal as possible. Different parameters in $\mu$ will increase the number of parameters in the light neutrino mass matrix ($M_\nu$). Even with the increased number of model parameters, it will be possible to fit with the neutrino oscillation data. However, to make the model more predictive, we have considered minimum number of parameters in the mass matrices involved in the model. In our work, $M_{D}$ is a ($3\times2$) matrix and contains six independent elements so that there are $^{6}\text{C}_{n}$ possibilities of texture zeros in $M_{D}$. The different zero textures of $M_{D}$ are discussed below. 

There are $^{6}\text{C}_{4} =15$ possible $4-0$ textures of Dirac mass matrix $M_{D}$. Again, the number of possible $3-0$ textures of Dirac mass matrix $M_{D}$ is  $^{6}\text{C}_{3} =20$. However, with the structures of $M_{N}$ and $\mu$, these give rise to $3-0$ and $2-0$ zero textures of $M_{\nu}$ which are not allowed. 

Now, the fifteen $2-0$ textures of $M_{D}$ in the framework of ISS $(2,3)$ are shown below:
\begin{table}[H]
	\begin{center}
		
		\scalebox{1.2}{
			\begin{tabular}{ |l| l| l|  }
				\hline
				\multicolumn{3}{|l|}{\hspace{23mm}2-0 textures of $M_{D}$} \\ 
				\hline
				$ M_{D}^{(1)}=\begin{pmatrix}0 & b \\
				c & 0 \\ 
				e & h  \end{pmatrix}$ &
				$ M_{D}^{(2)}=\begin{pmatrix}a & 0 \\
				0 & d \\ 
				e & h  \end{pmatrix}$&
				
				$M_{D}^{(3)}= \begin{pmatrix}a & b \\
				c & 0 \\ 
				0 & h \end{pmatrix}$\\
				\hline
				
			$M_{D}^{(4)}= \begin{pmatrix}a & b \\
			0 & d \\ 
			e & 0  \end{pmatrix}$ &
			$ M_{D}^{(5)}=\begin{pmatrix}0 & b \\
			c & d \\ 
			e & 0   \end{pmatrix}$&
			
			$ M_{D}^{(6)}=\begin{pmatrix}a & 0 \\
			c & d \\ 
			0 & h   \end{pmatrix}$\\
				\hline
				$ M_{D}^{(7)}=\begin{pmatrix}0 & b \\
				0 & d \\ 
				e & h  \end{pmatrix}$ &
				$ M_{D}^{(8)}=\begin{pmatrix}a & 0 \\
				c & 0 \\ 
				e & h   \end{pmatrix}$&
				
				$M_{D}^{(9)}= \begin{pmatrix}0 & b \\
				c & d \\ 
				0 & h  \end{pmatrix}$\\
				\hline
					$ M_{D}^{(10)}=\begin{pmatrix}a & 0 \\
					c & d \\ 
					e & 0   \end{pmatrix}$ &
				$M_{D}^{(11)}= \begin{pmatrix}a & b \\
				c & 0 \\ 
				e & 0  \end{pmatrix}$&
				
				$M_{D}^{(12)}= \begin{pmatrix}a & b \\
				0 & d \\ 
				0 & h  \end{pmatrix}$\\
				\hline
				$ M_{D}^{(13)}=\begin{pmatrix}0 & 0 \\
				c & d \\ 
				e & h  \end{pmatrix}$ &
				$ M_{D}^{(14)}=\begin{pmatrix}a & b \\
				0 & 0 \\ 
				e & h  \end{pmatrix}$&
				
				$ M_{D}^{(15)}=\begin{pmatrix}a & b \\
				c & d \\ 
				0 & 0  \end{pmatrix}$\\
				\hline
			\end{tabular}
		}
	\end{center}
	\caption{Possible two zero textures of $M_{D}$ in ISS(2,3)}
	\label{tabnew}
\end{table}

Among the structures of $M_{D}$ in table \ref{tabnew}, $M_{D}^{(13)}$,$M_{D}^{(14)}$ and $M_{D}^{(15)}$ along with $M_{N}$ and $\mu$ will lead to $M_{\nu}$ of the form given below:

\begin{equation} \label{eq:z} 
M_{\nu}= p\left(\begin{array}{ccc}
0 & 0 & 0\\
0 & \times & \times\\ 
0 & \times & \times
\end{array}\right),\;  M_{\nu}= p\left(\begin{array}{ccc}
\times & 0 & \times\\
0 & 0 & 0\\ 
\times & 0 & \times
\end{array}\right), \;  M_{\nu}= p\left(\begin{array}{ccc}
\times &\times & 0\\
\times & \times & 0\\ 
0 & 0 & 0
\end{array}\right)
\end{equation}

Here, $\times$ represents the non zero entry in $M_{\nu}$. The $M_{\nu}$ in Eq.\ref{eq:z} are discarded already in the literature. Thus the structures $M_{D}^{(13)}$,$M_{D}^{(14)}$ and $M_{D}^{(15)}$  are not allowed within the framework of ISS $(2,3)$. In our study, we have considered those structures of $M_{D}$ which lead to allowed light neutrino mass matrix.

The other 12 structures of $M_{D}$ given in table \ref{tabnew}, lead to different structures of $M_{\nu}$ which can be divided into six classes A1,A2,A3,B1,B2,B3.

Class A1:

\begin{equation} \label{eq:11} 
M_{D}^{(1)}= \left(\begin{array}{ccc}
0 & b \\
c & 0 \\ 
e & h 
\end{array}\right),\;  M_{D}^{(2)}= \left(\begin{array}{ccc}
a & 0 \\
0 & d \\ 
e & h 
\end{array}\right)
\end{equation}    
These two structures of $M_{D}$ lead to $1-0$ texture of  $M_{\nu}$ with zero at $(1,2)$ position.
\begin{equation} \label{eq:12} 
M_{\nu}= -p\left(\begin{array}{ccc}
k_{1} ^{2} & 0 & k_{1}k_{2} \\
0 & k_{2} ^{2} & k_{2}k_{3}\\ 
k_{1}k_{2} & k_{2}k_{3} & k_{1}^{2}+k_{4} ^{2}
\end{array}\right)
\end{equation}
where in $M_{D}^{(1)}$, $k_{1}= \frac{b}{g}$,$k_{2}= \frac{h}{g}$,$k_{3}= \frac{c}{f}$,$k_{4}= \frac{e}{f}$
and in $M_{D}^{(2)}$, $k_{1}= \frac{a}{g}$,$k_{2}= \frac{d}{g}$,$k_{3}= \frac{e}{g}$,$k_{4}= \frac{h}{g}$\\
Class A2:
\begin{equation} \label{eq:15} 
M_{D}^{(3)}= \left(\begin{array}{ccc}
a & b \\
c & 0 \\ 
0 & h 
\end{array}\right),\;  M_{D}^{(4)}= \left(\begin{array}{ccc}
a & b \\
0 & d \\ 
e & 0 
\end{array}\right)
\end{equation}    
These two structures of $M_{D}$ lead to $1-0$ texture of $M_{\nu}$ with zero at $(2,3)$ position.
\begin{equation} \label{eq:17} 
M_{\nu}= -p\left(\begin{array}{ccc}
k_{1} ^{2}+ k_{2}^{2} & k_{2}k_{3} & k_{1}k_{4}\\
k_{2}k_{3} & k_{3}^{2} & 0\\ 
k_{1}k_{4} & 0 & k_{4}^{2}
\end{array}\right)
\end{equation}
where in $M_{D}^{(3)}$, $k_{1}= \frac{b}{g}$,$k_{2}= \frac{a}{f}$, $k_{3}= \frac{c}{f}$,$k_{4}= \frac{h}{g}$
and in $M_{D}^{(4)}$, $k_{1}= \frac{a}{g}$,$k_{2}= \frac{b}{g}$,$k_{3}= \frac{d}{g}$,$k_{4}= \frac{e}{f}$\\

Class A3:
\begin{equation} \label{eq:13} 
 M_{D}^{(5)}= \left(\begin{array}{ccc}
0 & b \\
c & d \\ 
e & 0 
\end{array}\right),\;  M_{D}^{(6)}= \left(\begin{array}{ccc}
a & 0 \\
c & d \\ 
0 & h 
\end{array}\right)
\end{equation}    
These two structures of $M_{D}$ lead to $1-0$ texture of  $M_{\nu}$ with zero at $(1,3)$ position.
\begin{equation} \label{eq:14} 
M_{\nu}= -p\left(\begin{array}{ccc}
k_{1}^{2} &  k_{1} k_{3} & 0\\
k_{1} k_{3} & k_{1}^{2}+k_{3} ^{2} & k_{2} k_{4}\\ 
0 & k_{2} k_{4} & k_{4} ^{2} 
\end{array}\right)
\end{equation}
where in $M_{D}^{(5)}$, $k_{1}= \frac{b}{g}$,$k_{2}= \frac{d}{g}$, $k_{3}= \frac{c}{f}$,$k_{4}= \frac{e}{f}$
and in $M_{D}^{(6)}$, $k_{1}= \frac{a}{g}$,$k_{2}= \frac{d}{g}$,$k_{3}= \frac{c}{f}$,$k_{4}= \frac{h}{f}$\\
Class B1:
\begin{equation} \label{eq:21} 
 M_{D}^{(7)}= \left(\begin{array}{ccc}
0 & b \\
0 & d \\ 
e & h 
\end{array}\right),\;   M_{D}^{(8)}= \left(\begin{array}{ccc}
a & 0 \\
c & 0 \\ 
e & h 
\end{array}\right)
\end{equation}    
These two structures of $M_{D}$ lead to texture of  $M_{\nu}$ of the form:
\begin{equation} \label{eq:22} 
M_{\nu}= -p \left(\begin{array}{ccc}
k_{1} ^{2} & k_{1} k_{2} & k_{1} k_{3}\\
k_{1} k_{2} & k_{2} ^{2} & k_{2} k_{3}\\ 
k_{1} k_{3} &k_{2} k_{3} & k_{3} ^{2}+k_{4} ^{2}
\end{array}\right)
\end{equation}
where in $M_{D}^{(7)}$, $k_{1}= \frac{b}{g}$, $k_{2}= \frac{d}{g}$, $k_{3}= \frac{h}{g}$,$k_{4}= \frac{e}{f}$
and in $M_{D}^{(8)}$, $k_{1}= \frac{a}{f}$,$k_{2}= \frac{c}{f}$,$k_{3}= \frac{e}{f}$,$k_{4}= \frac{h}{g}$\\
Class B2:
\begin{equation} \label{eq:23} 
 M_{D}^{(9)}= \left(\begin{array}{ccc}
0 & b \\
c & d \\ 
0 & h 
\end{array}\right),\;   M_{D}^{(10)}= \left(\begin{array}{ccc}
a & 0 \\
c & d \\ 
e & 0 
\end{array}\right)
\end{equation}  
  
The structure of $M_{\nu}$ arising from these two structures of $M_{D}$ as follows:
\begin{equation} \label{eq:24} 
M_{\nu}= -p \left(\begin{array}{ccc}
k_{1} ^{2} & k_{1} k_{2} & k_{1} k_{3}\\
k_{1} k_{2} & k_{4} ^{2}+k_{2} ^{2} & k_{2} k_{3}\\ 
k_{1} k_{3} &k_{2} k_{3} & k_{3} ^{2}
\end{array}\right)
\end{equation}
where in $M_{D}^{(9)}$, $k_{1}= \frac{b}{g}$, $k_{2}= \frac{d}{g}$, $k_{3}= \frac{h}{g}$,$k_{4}= \frac{c}{f}$
and in $M_{D}^{(10)}$, $k_{1}= \frac{a}{f}$,$k_{2}= \frac{c}{f}$,$k_{3}= \frac{e}{f}$,$k_{4}= \frac{d}{g}$\\
Class B3:
\begin{equation} \label{eq:25} 
 M_{D}^{(11)}= \left(\begin{array}{ccc}
a & b \\
c & 0 \\ 
e & 0 
\end{array}\right),\;   M_{D}^{(12)}= \left(\begin{array}{ccc}
a & b \\
0 & d \\ 
0 & h 
\end{array}\right)
\end{equation}    
These two structures of $M_{D}$ give rise to the following texture of $M_{\nu}$:
\begin{equation} \label{eq:26} 
M_{\nu}= -p \left(\begin{array}{ccc}
k_{1} ^{2}+k_{2} ^{2} & k_{2} k_{3} & k_{2} k_{4}\\
k_{2} k_{3} & k_{3}^{2} & k_{3} k_{4}\\ 
k_{2} k_{4} &k_{3} k_{4} & k_{4} ^{2}
\end{array}\right)
\end{equation}
where in $M_{D}^{(11)}$, $k_{1}= \frac{b}{g}$, $k_{2}= \frac{a}{f}$, $k_{3}= \frac{c}{f}$,$k_{4}= \frac{e}{f}$
and in $M_{D}^{(12)}$, $k_{1}= \frac{a}{f}$,$k_{2}= \frac{b}{g}$,$k_{3}= \frac{d}{g}$,$k_{4}= \frac{h}{g}$\\

As mentioned earlier, ISS$(2,3)$ leads to a sterile state in the keV scale and we have studied the phenomenology of keV scale sterile neutrino in our earlier work. However, we have studied a particular $S_{4}$ flavor symmetric model. In this paper, we study the impact of different texture zero of the mass matrices on the dark matter phenomenology. We carry our study with those structures of $M_{\nu}$ which are already allowed in the literature.
\section{\label{sec:level4}Sterile neutrino dark matter in ISS (2,3)}
As mentioned above, ISS$(2,3)$ can naturally lead to a sterile state of several keV
and with very small mixing with the active neutrinos. Thus this model provides a particle having a lifetime exceeding the age of the Universe, which can be considered as a warm dark matter candidate \cite{Dolgov:2000ew,Kusenko:2009up,Abazajian:2017tcc,Dodelson:1993je,Kusenko:2009up,Merle:2017jfn}. Again, texture zeros of mass matrices can have a significant impact on dark matter phenomenology as the mass and mixing are strongly dependent on the textures which in turn constrain the other DM parameters. There are various constraints and bounds on sterile neutrino dark matter coming from experiments as well as some cosmological considerations \cite{Bezrukov:2009th,Campos:2016gjh,Horiuchi:2013noa}. To study DM phenomenology, three main aspects are to be considered, relic abundance, stability (lifetime),and structure formation \cite{Boyarsky:2018tvu}.

The relic abundance of sterile neutrino within the framework of ISS$(2,3)$ can be expressed as \cite{Abada:2014zra}:
\begin{equation}\label{eq:c}
\Omega_{DM}h^{2} = 1.1 \times 10^{7}\sum C_{\alpha}(m_{s})|\mathcal{U}_{\alpha s}|^{2}{\left(\frac{m_{s}}{keV}\right)}^{2},  \alpha = e,\mu,\tau
\end{equation}
 It can be simplified to the following expression,
\begin{equation}
\Omega_{DM}h^{2}\approxeq0.3 {\left(\frac{sin^{2}2\theta}{10^{-10}}\right)}{\left(\frac{m_{s}}{100keV}\right)}^{2}
\end{equation}
where $sin^{2}2\theta = 4\sum|\mathcal{U}_{\alpha s}|^{2}$ with $|\mathcal{U}_{\alpha s}|$ being the active-sterile leptonic mixing matrix element and $m_{s}$ represents the mass of the lightest sterile fermion.

Inverse seesaw provides a decaying DM as the sterile neutrino is not completely stable. A sub dominant fraction of sterile neutrino decays into an active neutrino and a photon $\gamma$ via the process $N\longrightarrow\nu+\gamma$ that produces a narrow line in the x-ray spectrum of astrophysical objects \cite{Dolgov:2000ew,Abazajian:2001vt}. This poses strong bounds on the active-sterile mixing and the mass of the sterile neutrino \cite{Ng:2019gch,Boyarsky:2007ay}. To be a viable DM candidate, the lifetime of the decaying sterile neutrino should be greater than the age of the Universe, which restricts the decay width of the process. The decay rate can be written as \cite{Ng:2019gch} :
\begin{equation}\label{eq:d}
\Gamma=1.38\times10^{-32}{\left(\frac{sin^{2}2\theta}{10^{-10}}\right)}{\left(\frac{m_{s}}{keV}\right)}^{5}s^{-1}.\end{equation}

An important constraint on sterile neutrino dark matter is Lyman-$\alpha$ bound \cite{Seljak:2006qw,Boyarsky:2008xj}. Ly-$\alpha$ provides stronger bounds on the velocity distribution of the DM particles from the effect of their free streaming on the large scale structure formation \cite{Boyarsky:2008mt}. This constraint can be converted into a bound for the mass of the DM particle which can be seen in \cite{Boyarsky:2008mt}. The constraint is strongly model dependent and the bounds are governed by the production mechanism of the DM candidate. In this work, we have adopted the bounds considering XQ-100 Ly-$\alpha$ data given in \cite{Baur:2017stq,Boyarsky:2008xj}. 

\section{\label{sec:level5} Neutrinoless Double Beta Decay (0$\nu\beta\beta$)}
Apart from the SM particles, ISS (2,3) contains eight extra heavy states, the presence of which may lead to many new contributions to lepton number violating processes like neutrinoless double beta decay(0$\nu\beta\beta$)\cite{Benes:2005hn,Awasthi:2013we,Borgohain:2018lro}. We have studied the potential contributions of the heavy states to the effective electron neutrino Majorana mass $m_{ee}$ \cite{Abada:2018qok,Blennow:2010th} characterising 0$\nu\beta\beta$. KamLAND-ZEN, GERDA, CUORE and EX0-200 provide stringent bounds on $m_{ee}$ which can be seen in \cite{KamLAND-Zen:2016pfg,Adams:2019jhp,Anton:2019wmi,Agostini_2020}.

The decay width of the process is proportional to the effective electron neutrino Majorana mass $m_{ee}$. In the absence of any sterile neutrino,the standard contribution to $m_{ee}$ can be written as,
\begin{equation}
m_{ee} = \mathrel{\Big|}\sum_{i = 1}^{3}{U_{ei}}^{2}m_{i}\mathrel{\Big|}
\end{equation}
The above equation is modified because of the presence of the heavy neutrinos and is given by \cite{Abada:2018qok} 
\begin{equation}\label{eq:30}
m_{ee}  =\mathrel{\Big|}\sum_{i = 1}^{3}{U_{ei}}^{2}m_{i}\mathrel{\Big|} + \mathrel{\Big|}\sum_{j = 1}^{5}{U_{ej}}^{2}\frac{M_{j}}{k^{2}+M_{j}^{2}}|<k>|^{2}\mathrel{\Big|}
\end{equation}
where, ${U_{ej}}$ represents the coupling of the heavy neutrinos to the electron neutrino and $M_{j}$ represents the mass of the respective heavy neutrinos. $|<k>|$ is known as neutrino virtuality momentum with value $|<k>|\simeq 190$ MeV.
\section{\label{sec:level6} Results of Numerical Analaysis and Discussions}
We analyse light neutrino mass matrix $M_{\nu}$ arising from 15 two zero textures of Dirac neutrino mass matrix $M_{D}$ with maximal zeros in $\mu$ and $M_{N}$ involved in this framework. Out of 15 two zero textures of $M_{D}$, three lead to $M_{\nu}$ which do not give correct neutrino phenomenology. The remaining 12 textures can be classified into six  different categories. The model parameters of the remaining categories have been evaluated using 
\begin{equation}\label{eq:16}
M_{\nu} = U_{\text{PMNS}}M^{\text{diag}}_{\nu} U^T_{\text{PMNS}}
\end{equation}
where the Pontecorvo-Maki-Nakagawa-Sakata (PMNS) leptonic mixing matrix can be parametrized as \cite{Giunti:2007ry}
\begin{equation}
U_{\text{PMNS}}=\left(\begin{array}{ccc}
c_{12}c_{13}& s_{12}c_{13}& s_{13}e^{-i\delta}\\
-s_{12}c_{23}-c_{12}s_{23}s_{13}e^{i\delta}& c_{12}c_{23}-s_{12}s_{23}s_{13}e^{i\delta} & s_{23}c_{13} \\
s_{12}s_{23}-c_{12}c_{23}s_{13}e^{i\delta} & -c_{12}s_{23}-s_{12}c_{23}s_{13}e^{i\delta}& c_{23}c_{13}
\end{array}\right) U_{\text{Maj}}
\label{matrixPMNS}
\end{equation}
where $c_{ij} = \cos{\theta_{ij}}, \; s_{ij} = \sin{\theta_{ij}}$ and $\delta$ is the leptonic Dirac CP phase. The diagonal matrix $U_{\text{Maj}}=\text{diag}(1, e^{i\alpha}, e^{i(\beta+\delta)})$  contains the Majorana CP phases $\alpha, \beta$.
The diagonal mass matrix of the light neutrinos can be written  as, $M^{\text{diag}}_{\nu} 
= \text{diag}(0, \sqrt{m^2_1+\Delta m_{solar}^2}, \sqrt{m_1^2+\Delta m_{atm}^2})$ for normal oredring (NO) and  $M^{\text{diag}}_{\nu} = \text{diag}(\sqrt{m_3^2+\Delta m_{atm}^2}, 
\sqrt{\Delta m_{solar}^2+ \Delta m_{atm}^2}, m_3)$ for inverted ordering (IO) \cite{Nath:2016mts}.
We use neutrino oscillation parameters in $3\sigma$ range as given in table \ref{tab3} as input and constrain our parameter space. 

After evaluating the model parameters in all categories, we calculate the mass of the sterile neutrino dark matter $m_{DM}$ and DM-active mixing using Eq.\eqref{eq:1}. Then we obtain the relic abundance of the proposed DM using Eq.\eqref{eq:c}. Again, with the mass and mixing of the dark matter, we evaluate the decay rates of the sterile DM in each of the above categories.

\begin{figure}[H]
	\begin{center}
		\includegraphics[width=0.30\textwidth]{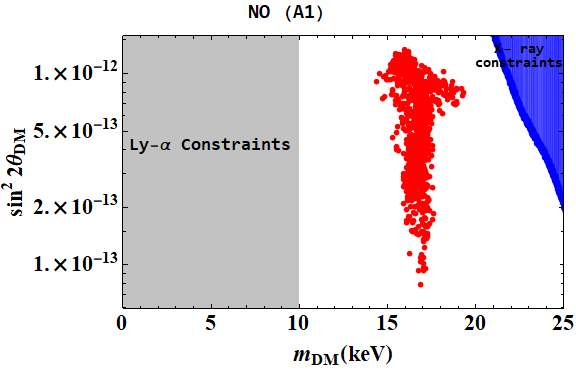}
		\includegraphics[width=0.30\textwidth]{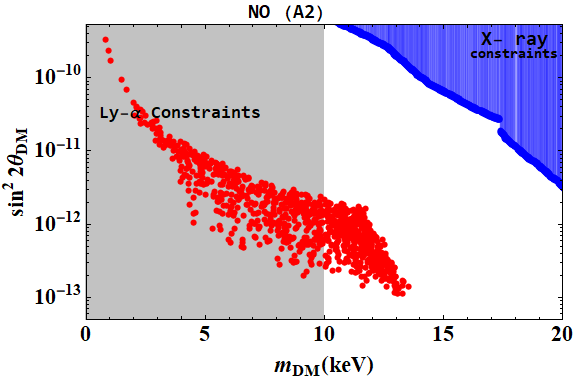}
		\includegraphics[width=0.30\textwidth]{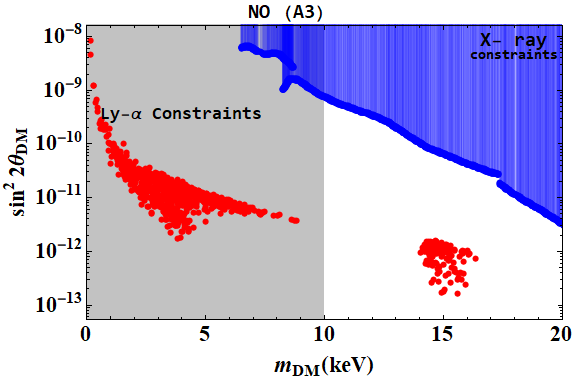}\\
		\includegraphics[width=0.30\textwidth]{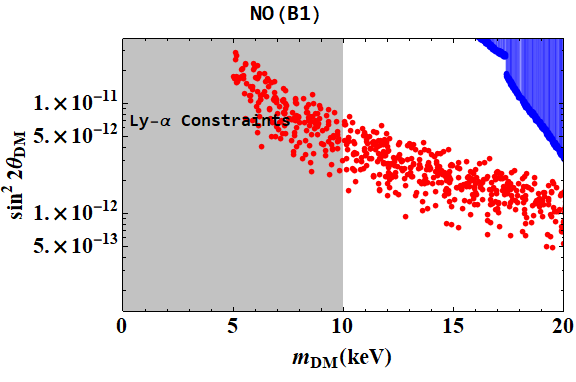}
		\includegraphics[width=0.30\textwidth]{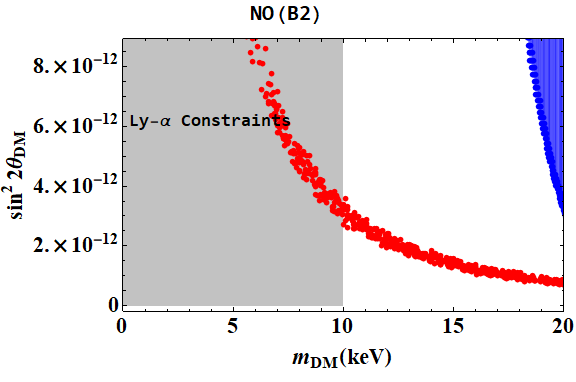}
		\includegraphics[width=0.30\textwidth]{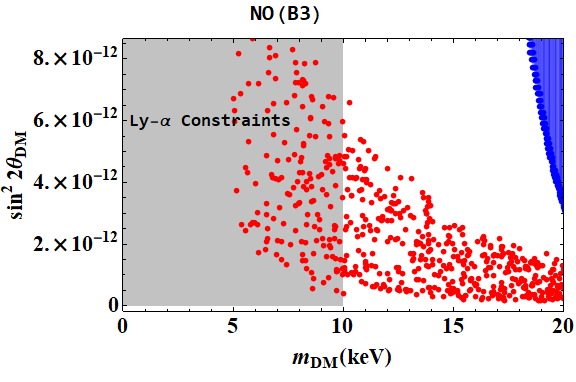}
	\end{center}
	\begin{center}
		\caption{DM-active mixing as a function of the mass of the DM for NO.The gray region represents the Ly-$\alpha$ bound and the blue region represents the X-ray bound.}
		\label{fig1}
	\end{center}
\end{figure}

Fig \ref{fig1} to fig \ref{fig8} represent the results obtained from the numerical analysis for all the allowed textures within the framework of ISS(2,3). 
\begin{figure}[H]
	\begin{center}
		\includegraphics[width=0.30\textwidth]{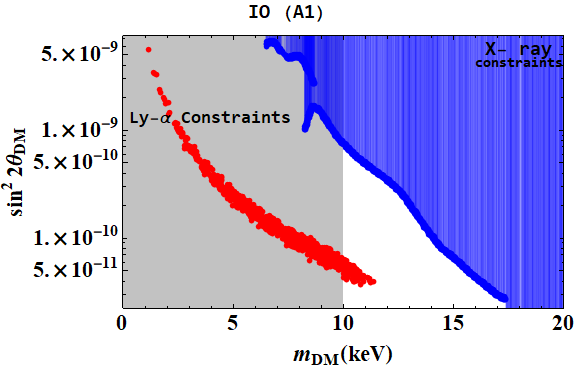}
		\includegraphics[width=0.30\textwidth]{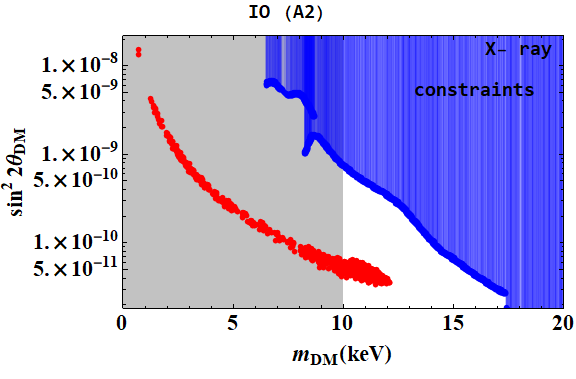}
		\includegraphics[width=0.30\textwidth]{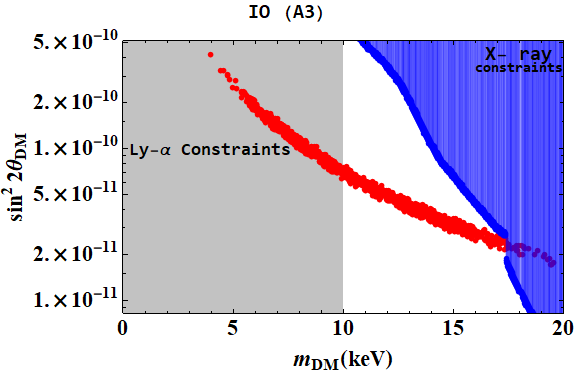}\\
		\includegraphics[width=0.30\textwidth]{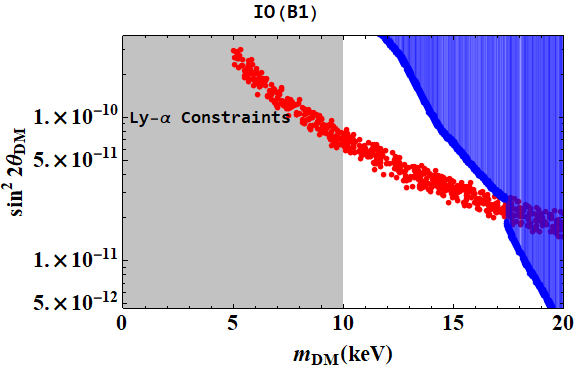}
		\includegraphics[width=0.30\textwidth]{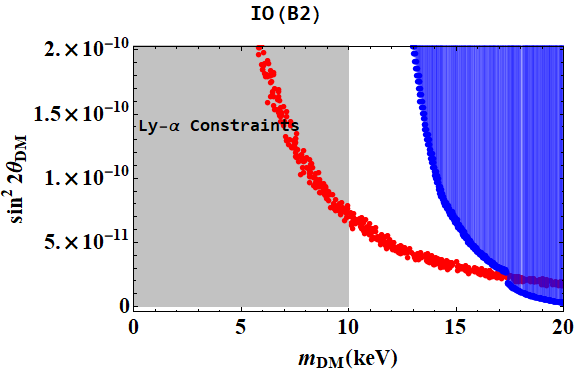}
		\includegraphics[width=0.30\textwidth]{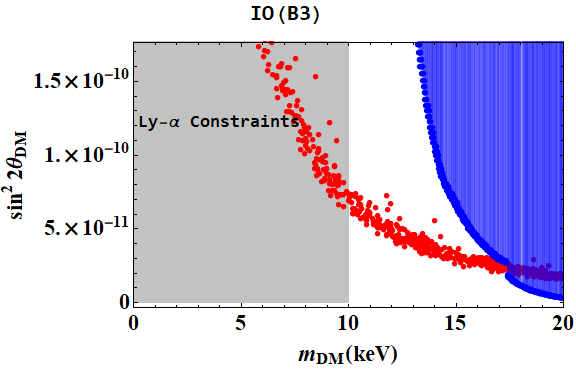}
	\end{center}
	\begin{center}
		\caption{DM-active mixing as a function of the mass of the DM for IO in different textures.The gray region represents the Ly-$\alpha$ bound and the blue region represents the X-ray bound.}
		\label{fig1a}
	\end{center}
\end{figure} 
\begin{figure}[H]
	\begin{center}
		\includegraphics[width=0.30\textwidth]{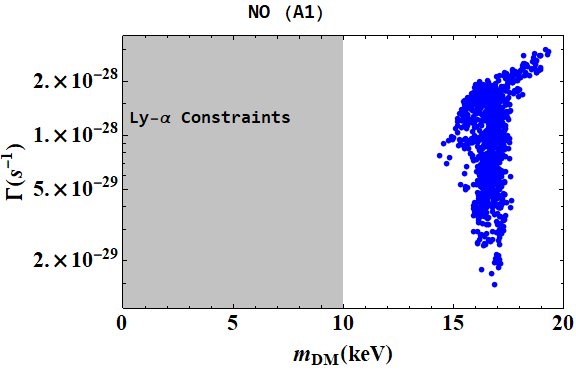}
		\includegraphics[width=0.30\textwidth]{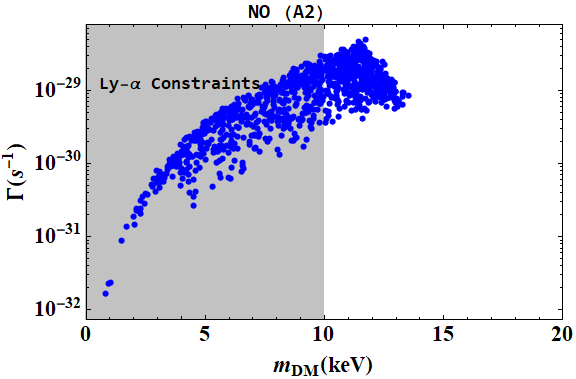}
		\includegraphics[width=0.30\textwidth]{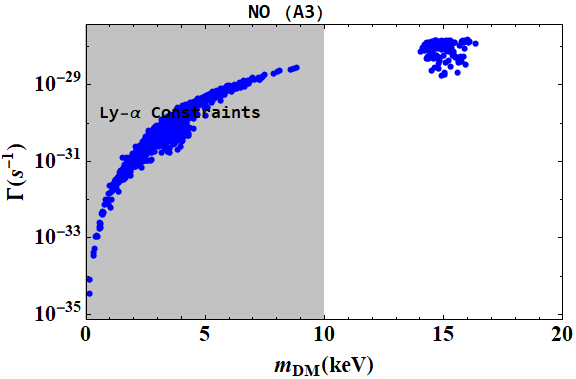}\\
		\includegraphics[width=0.30\textwidth]{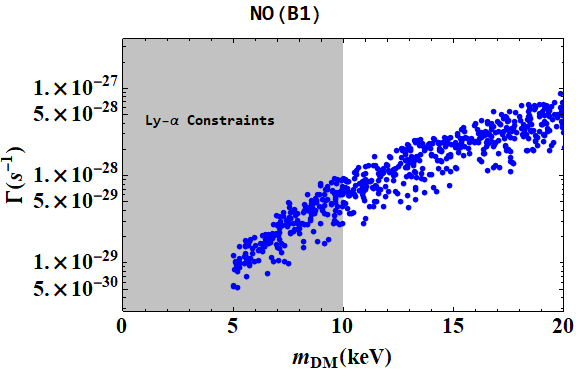}
		\includegraphics[width=0.30\textwidth]{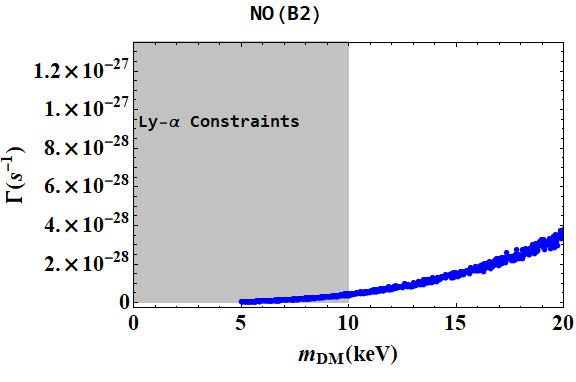}
		\includegraphics[width=0.30\textwidth]{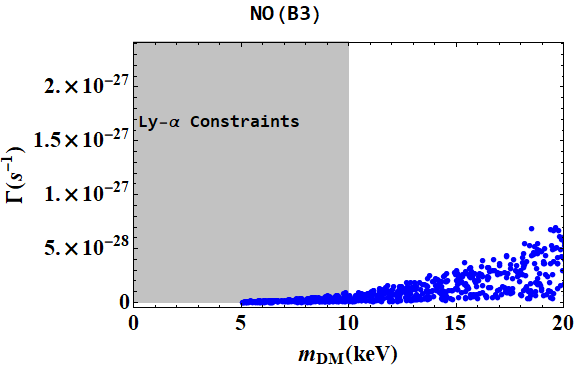}
	\end{center}
	\begin{center}
		\caption{Decay rate (in $s^{-1}$)of the lightest sterile neutrino as a function of DM mass for NO in different textures.}
		\label{fig2}
	\end{center}
\end{figure} 
\begin{figure}[H]
	\begin{center}
		\includegraphics[width=0.30\textwidth]{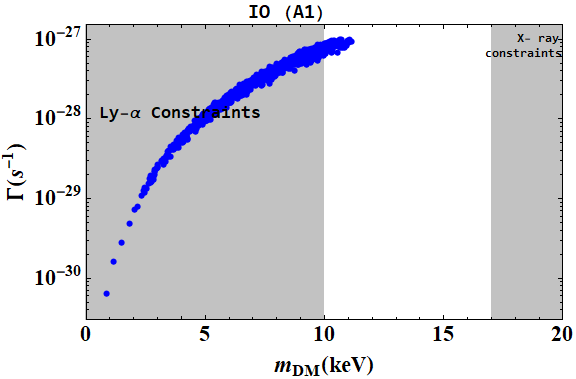}
		\includegraphics[width=0.30\textwidth]{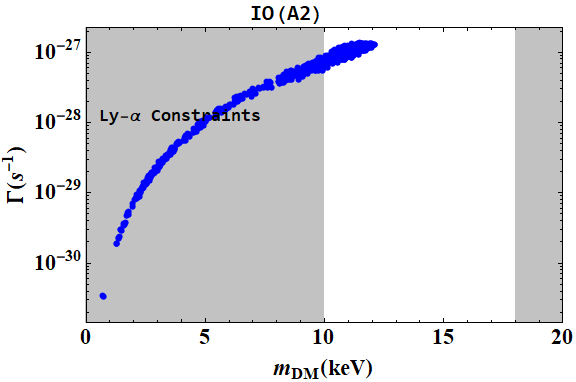}
		\includegraphics[width=0.30\textwidth]{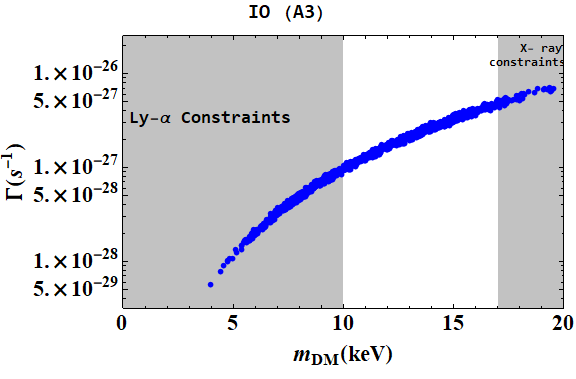}\\
		\includegraphics[width=0.30\textwidth]{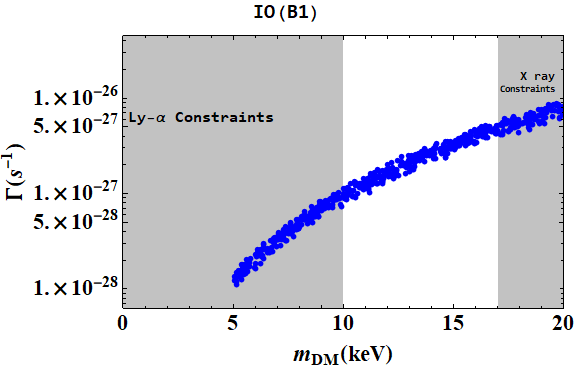}
		\includegraphics[width=0.30\textwidth]{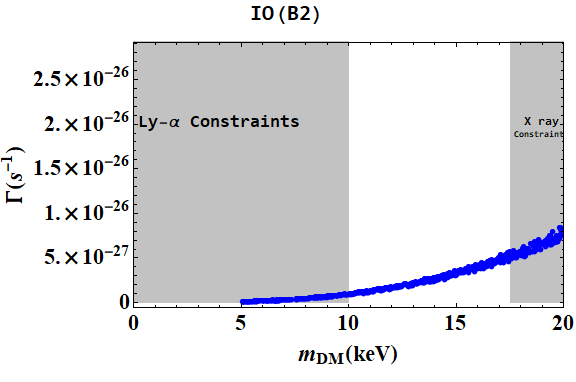}
		\includegraphics[width=0.30\textwidth]{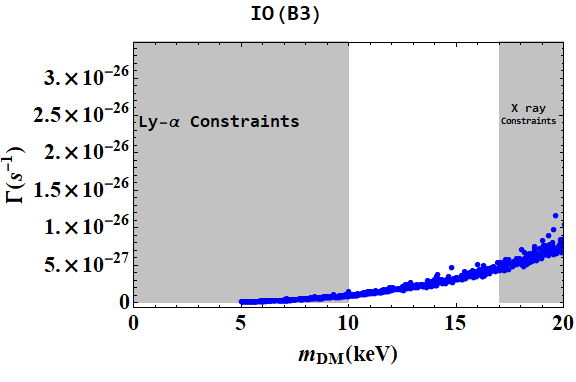}
	\end{center}
	\begin{center}
		\caption{Decay rate (in $s^{-1}$)of the lightest sterile neutrino as a function of DM mass in six different textures in case of IO.}
		\label{fig2a}
	\end{center}
\end{figure} 
\begin{figure}[H]
	\begin{center}
		\includegraphics[width=0.30\textwidth]{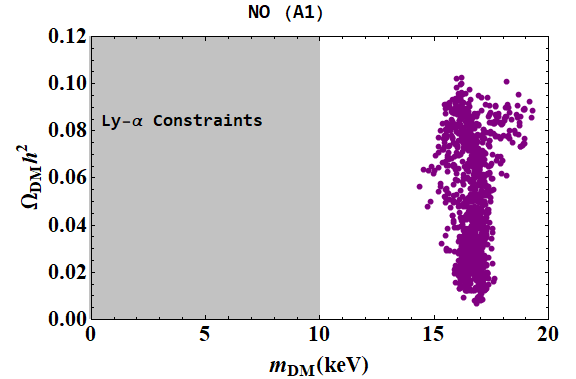}
		\includegraphics[width=0.30\textwidth]{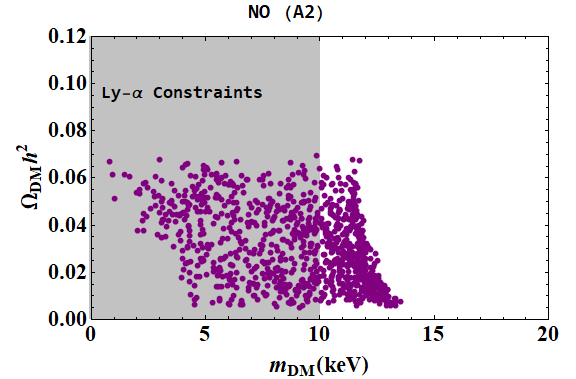}
		\includegraphics[width=0.30\textwidth]{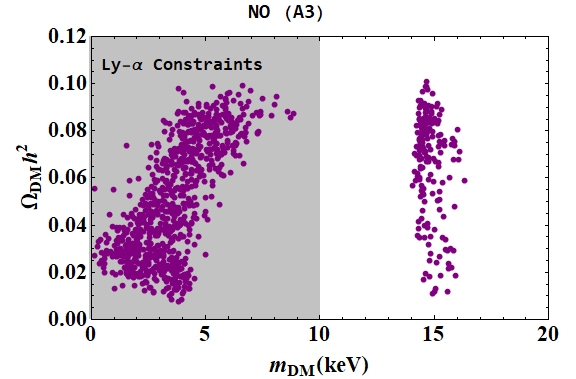}\\
		\includegraphics[width=0.30\textwidth]{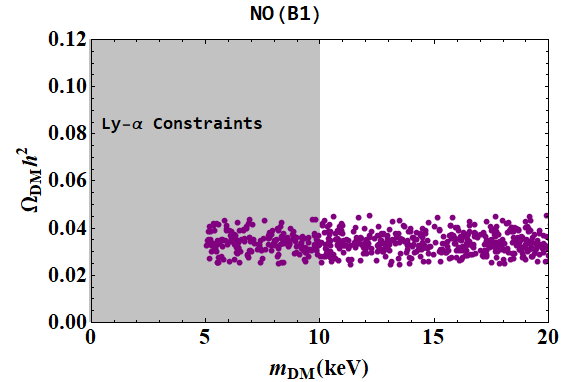}
		\includegraphics[width=0.30\textwidth]{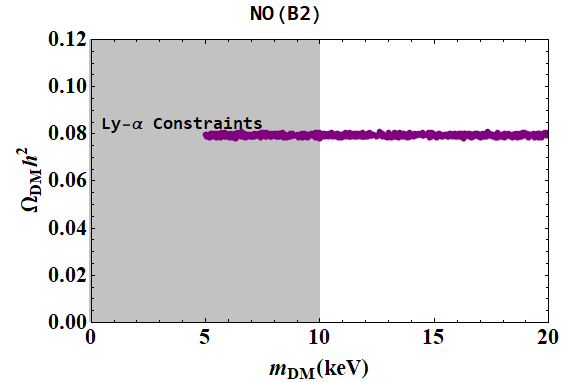}
		\includegraphics[width=0.30\textwidth]{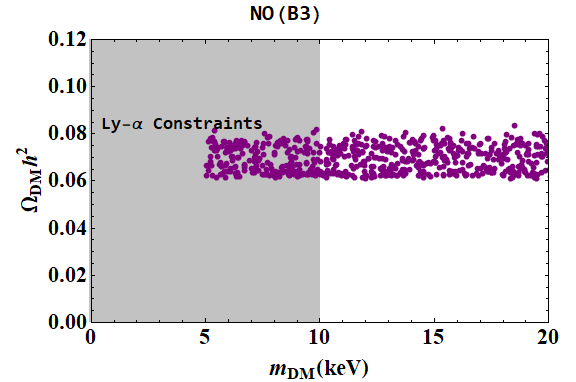}
	\end{center}
	\begin{center}
		\caption{Predictions of different textures on relic abundance of sterile neutrino dark matter in NO.}
		\label{fig3}
	\end{center}
\end{figure} 
\begin{figure}[H]
	\begin{center}
		\includegraphics[width=0.30\textwidth]{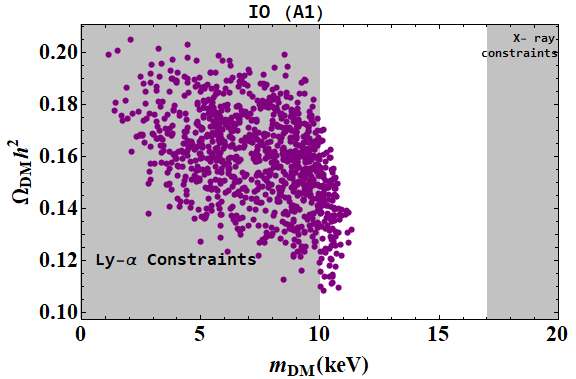}
		\includegraphics[width=0.30\textwidth]{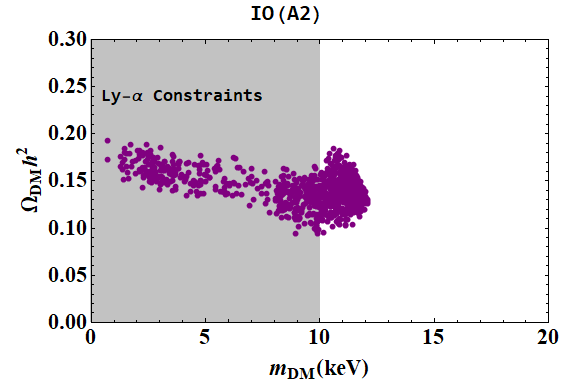}
		\includegraphics[width=0.30\textwidth]{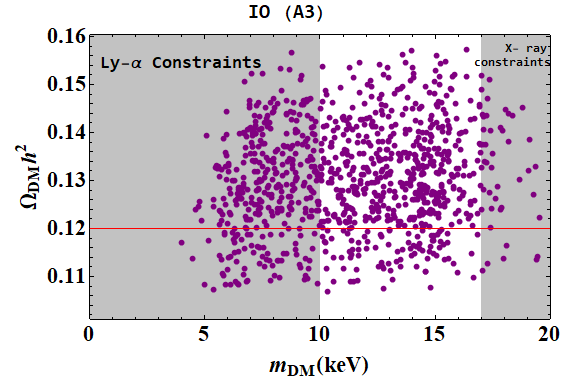}\\
		\includegraphics[width=0.30\textwidth]{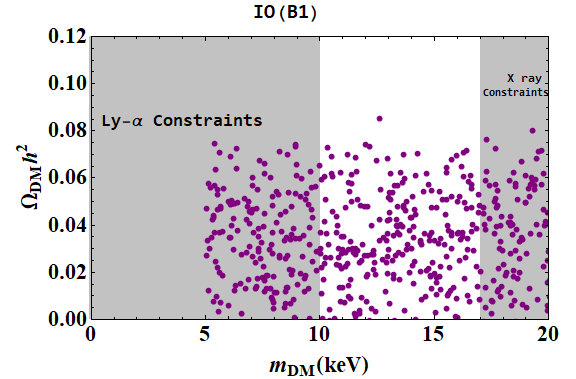}
		\includegraphics[width=0.30\textwidth]{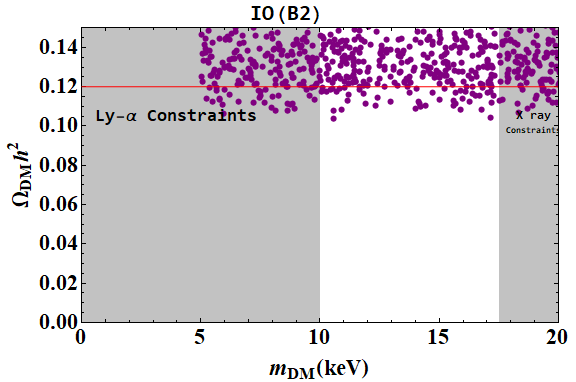}
		\includegraphics[width=0.30\textwidth]{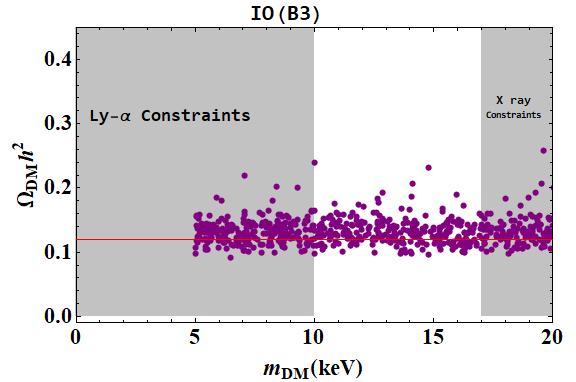}
	\end{center}
	\begin{center}
		\caption{Predictions of different textures on relic abundance of sterile neutrino dark matter in IO.Red horizontal line in the plots represents the present dark matter abundance.}
		\label{fig3a}
	\end{center}
\end{figure} 
\begin{figure}[H]
	\begin{center}
		\includegraphics[width=0.30\textwidth]{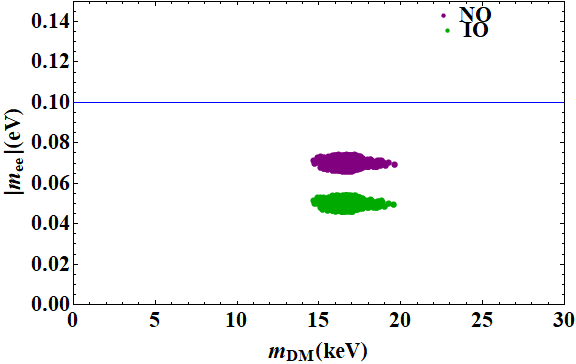}
		\includegraphics[width=0.30\textwidth]{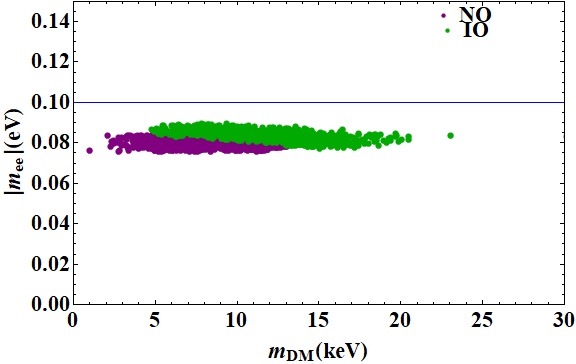}
		\includegraphics[width=0.30\textwidth]{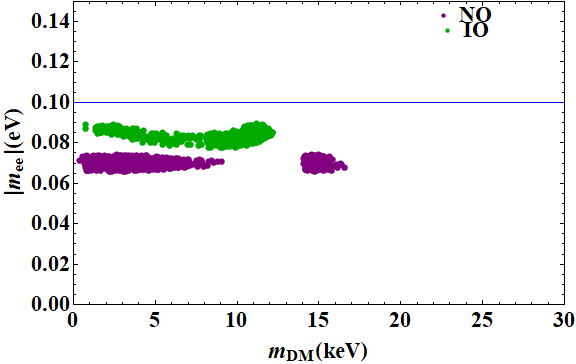}\\
		\includegraphics[width=0.30\textwidth]{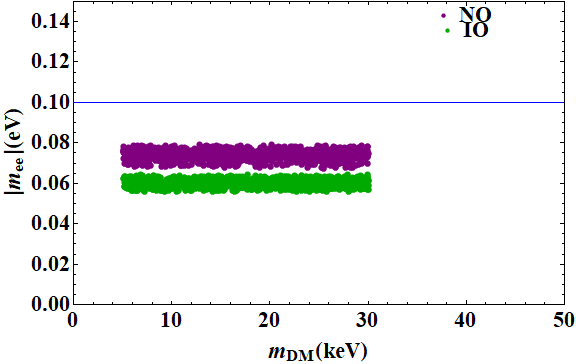}
		\includegraphics[width=0.30\textwidth]{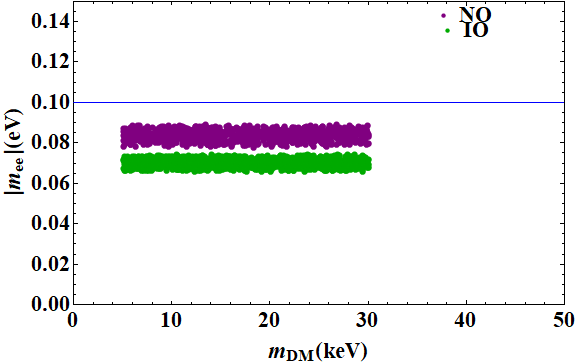}
		\includegraphics[width=0.30\textwidth]{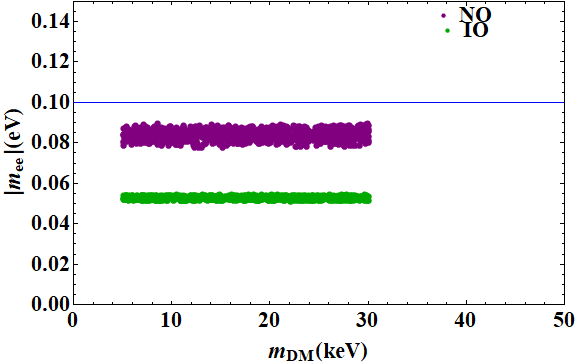}
	\end{center}
	\begin{center}
		\caption{Effective mass as a function of mass of DM in different textures for both NO and IO.}
		\label{fig4}
	\end{center}
\end{figure}
\begin{figure}[H]
	\begin{center}
		\includegraphics[width=0.30\textwidth]{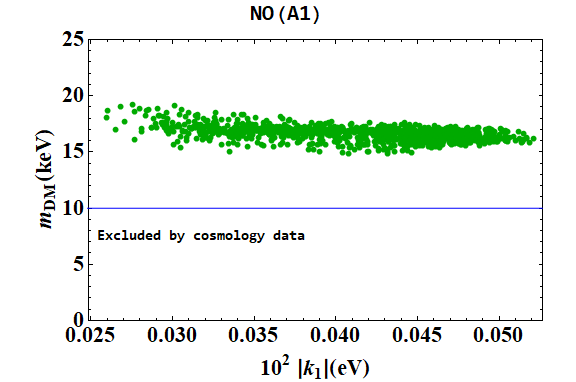}
		\includegraphics[width=0.30\textwidth]{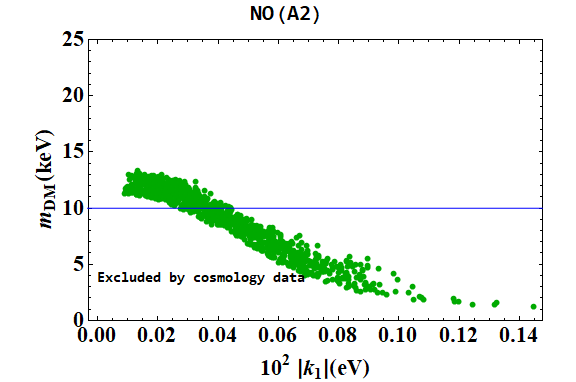}
		\includegraphics[width=0.30\textwidth]{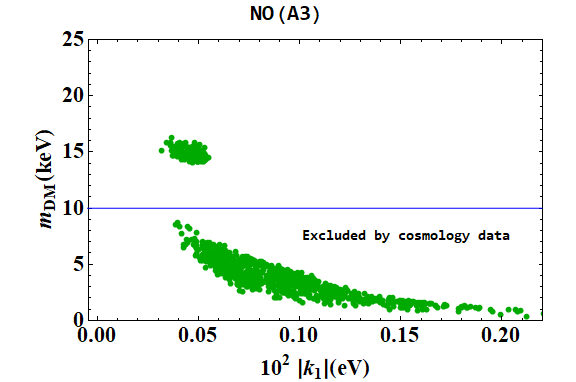}\\
		\includegraphics[width=0.30\textwidth]{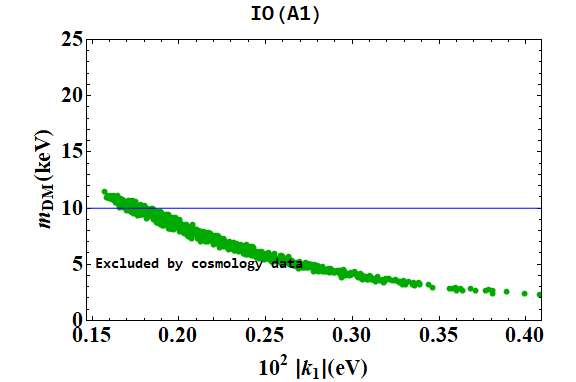}
		\includegraphics[width=0.30\textwidth]{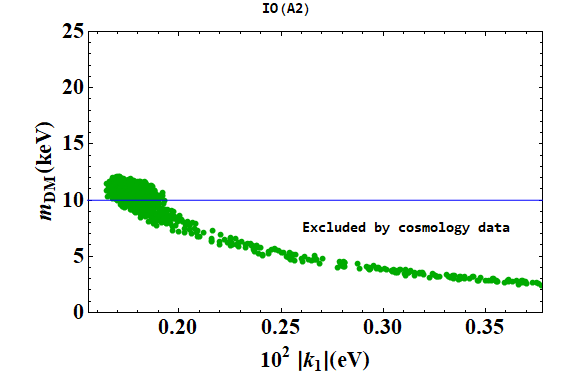}
		\includegraphics[width=0.30\textwidth]{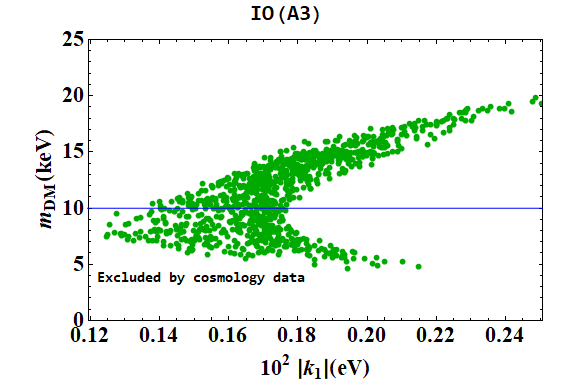}\\
        \includegraphics[width=0.30\textwidth]{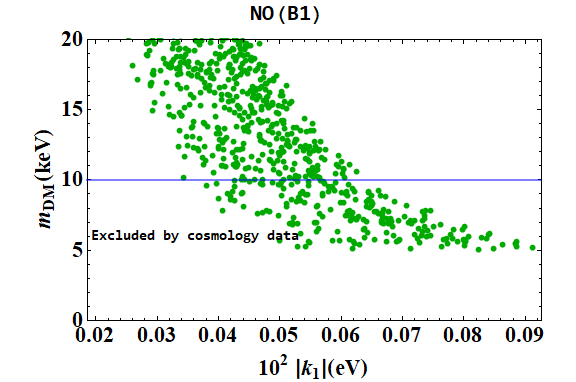}
		\includegraphics[width=0.30\textwidth]{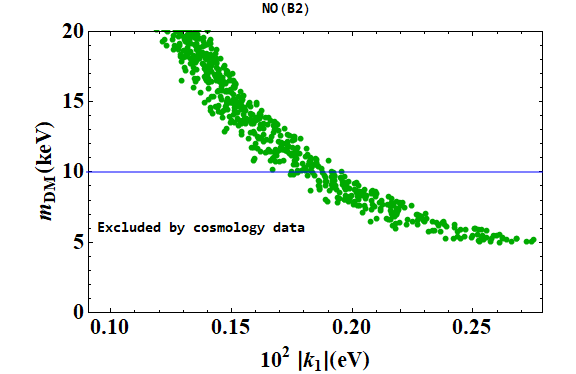}
		\includegraphics[width=0.30\textwidth]{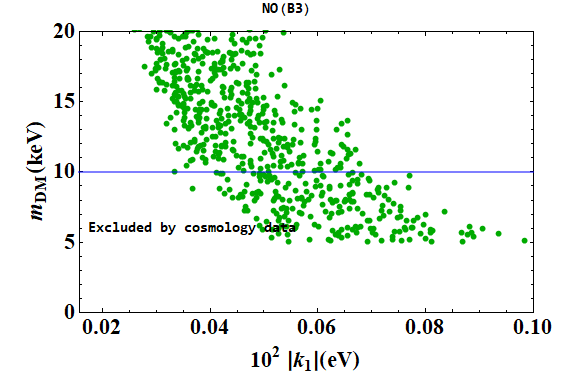}\\
	    \includegraphics[width=0.30\textwidth]{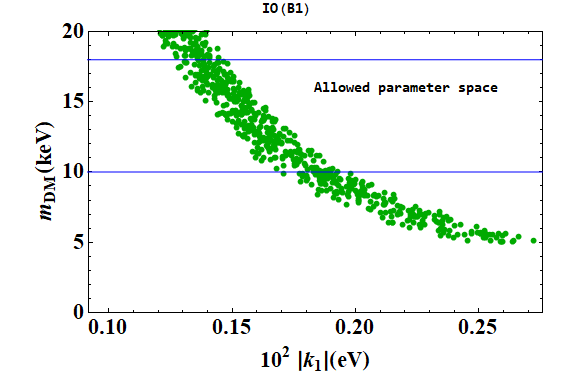}
	    \includegraphics[width=0.30\textwidth]{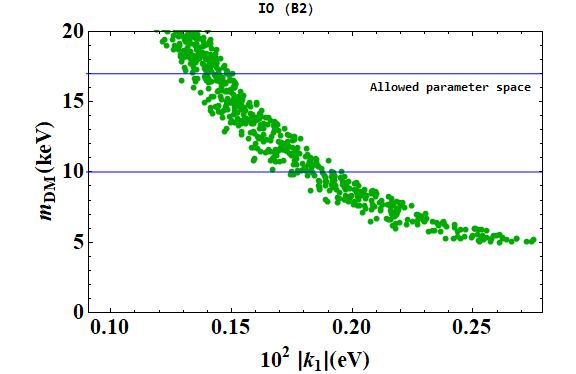}
  	    \includegraphics[width=0.30\textwidth]{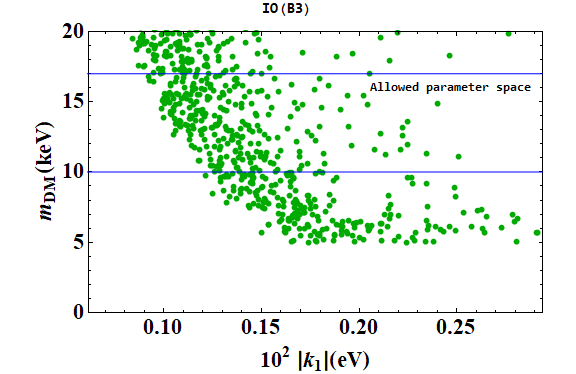}
\end{center}
	\begin{center}
		\caption{Dark matter mass as a function of model parameter k1 for NO as well as IO}
		\label{fig5}
	\end{center}
\end{figure}

\begin{figure}[H]
	\begin{center}
		\includegraphics[width=0.30\textwidth]{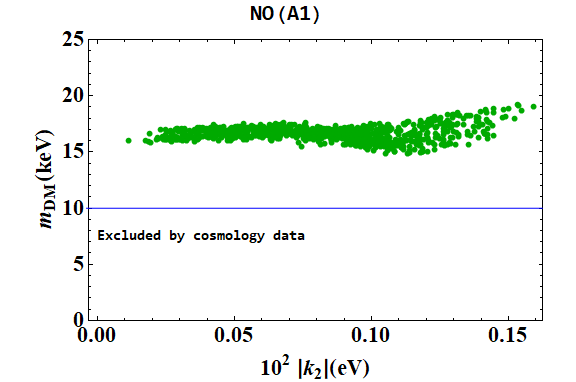}
		\includegraphics[width=0.30\textwidth]{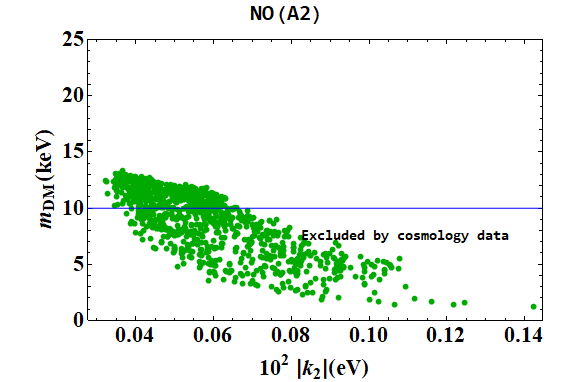}
		\includegraphics[width=0.30\textwidth]{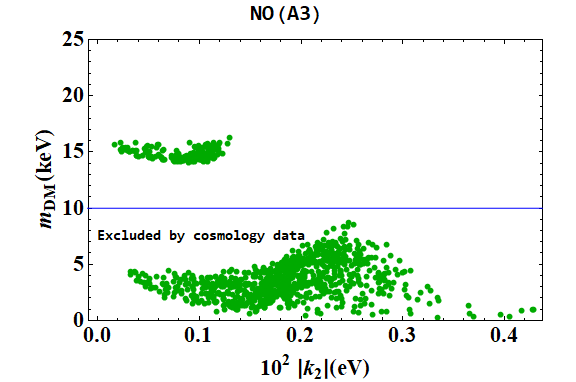}\\
		\includegraphics[width=0.30\textwidth]{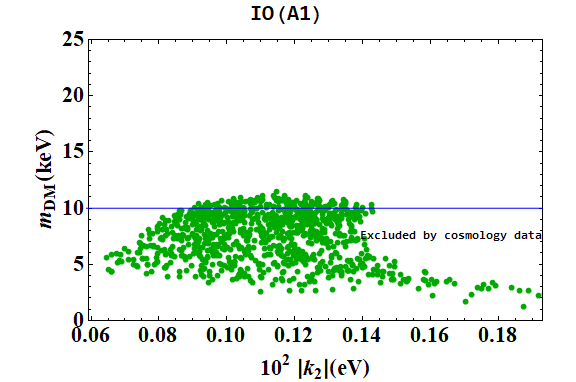}
		\includegraphics[width=0.30\textwidth]{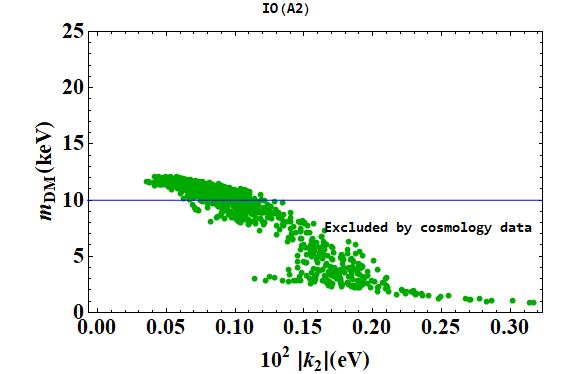}
		\includegraphics[width=0.30\textwidth]{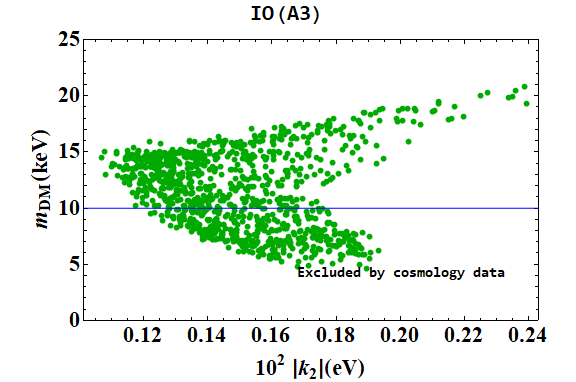}\\
		\includegraphics[width=0.30\textwidth]{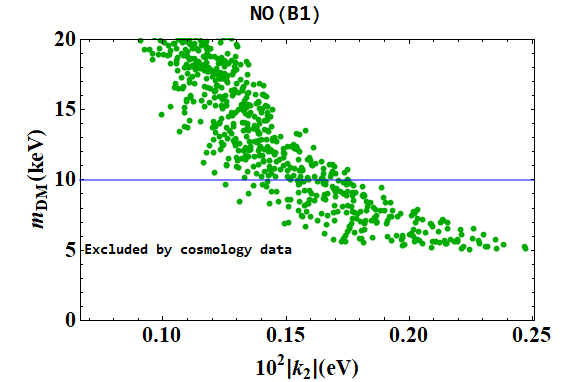}
		\includegraphics[width=0.30\textwidth]{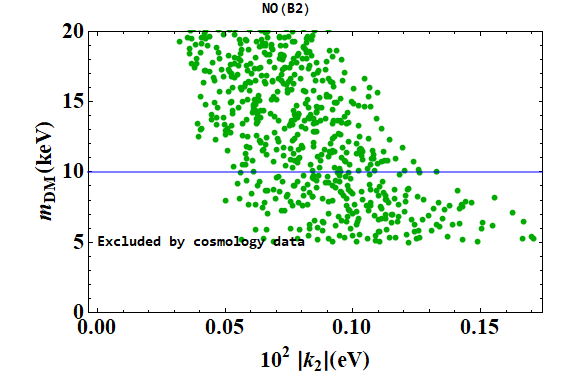}
		\includegraphics[width=0.30\textwidth]{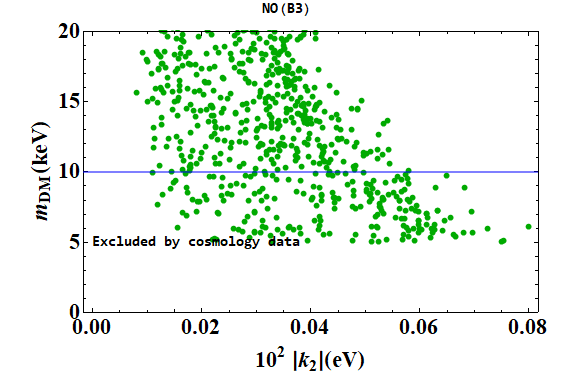}\\
		\includegraphics[width=0.30\textwidth]{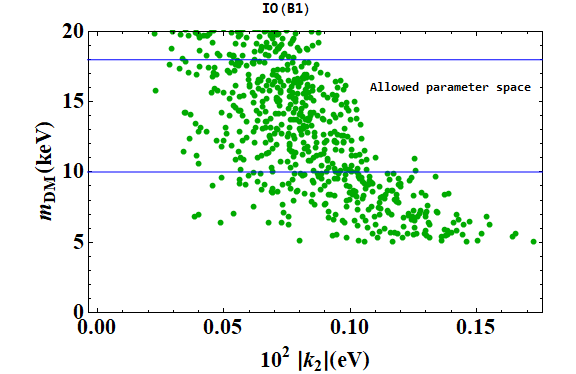}
		\includegraphics[width=0.30\textwidth]{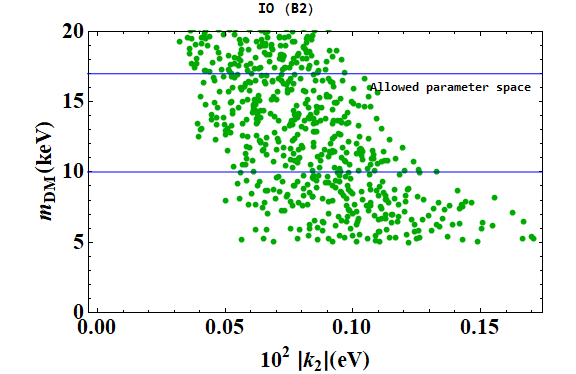}
		\includegraphics[width=0.30\textwidth]{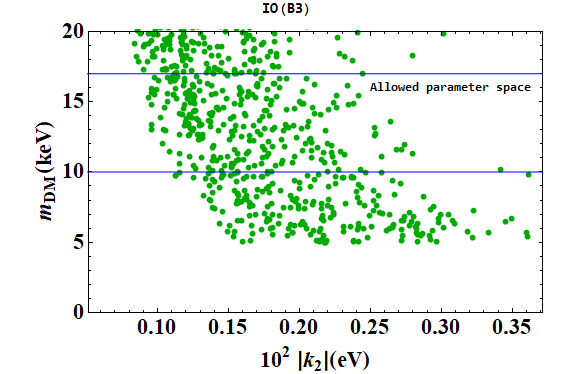}
	\end{center}
	\begin{center}
		\caption{Dark matter mass as a function of model parameter k2 for NO as well as IO}
		\label{fig6}
	\end{center}
\end{figure}
\begin{figure}[H]
	\begin{center}
		\includegraphics[width=0.30\textwidth]{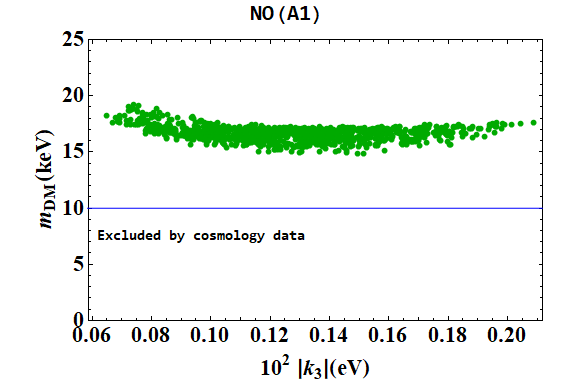}
		\includegraphics[width=0.30\textwidth]{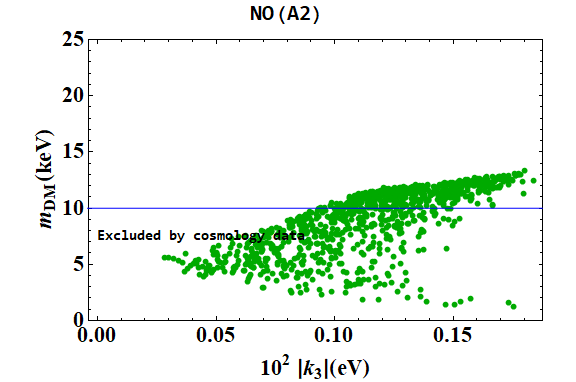}
		\includegraphics[width=0.30\textwidth]{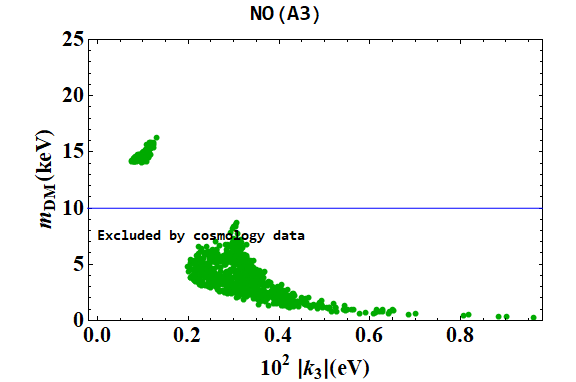}\\
		\includegraphics[width=0.30\textwidth]{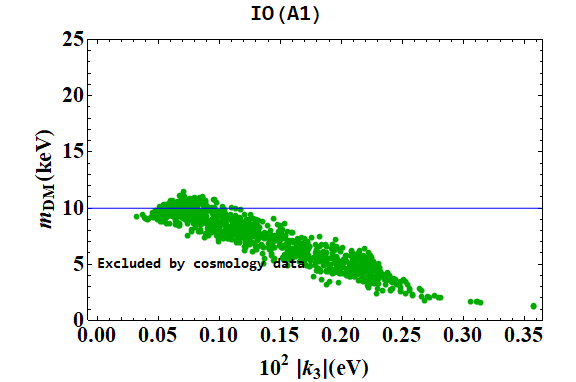}
		\includegraphics[width=0.30\textwidth]{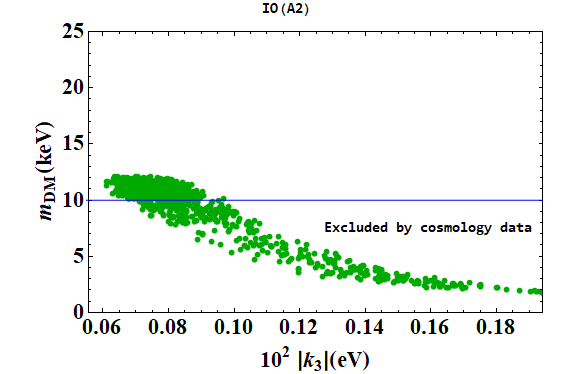}
		\includegraphics[width=0.30\textwidth]{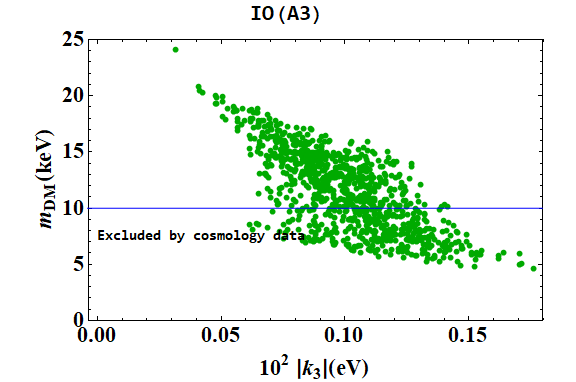}\\
		\includegraphics[width=0.30\textwidth]{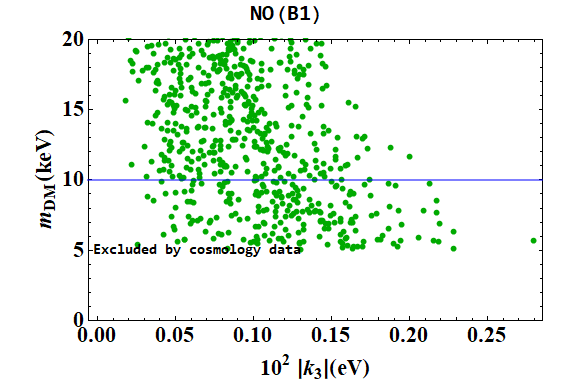}
		\includegraphics[width=0.30\textwidth]{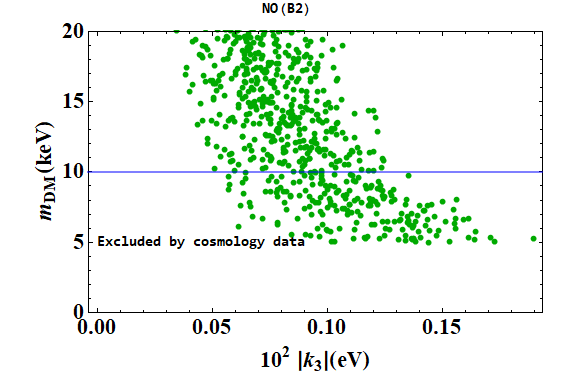}
		\includegraphics[width=0.30\textwidth]{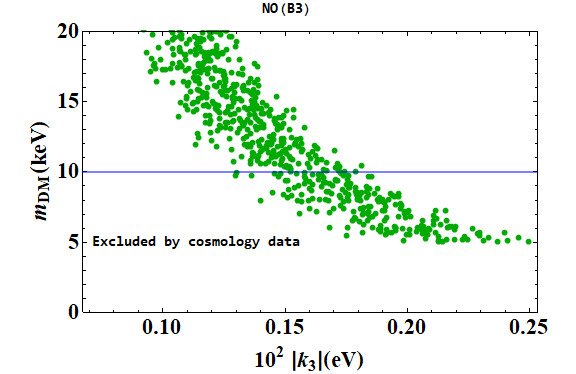}\\
		\includegraphics[width=0.30\textwidth]{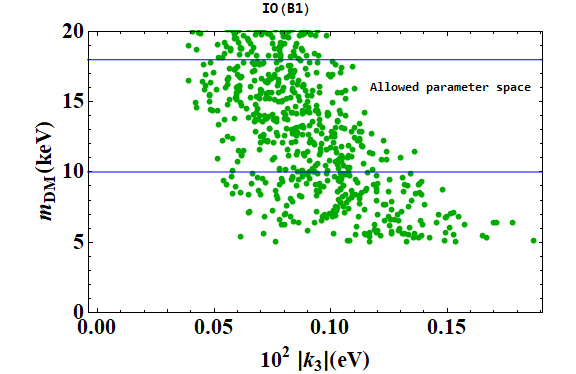}
		\includegraphics[width=0.30\textwidth]{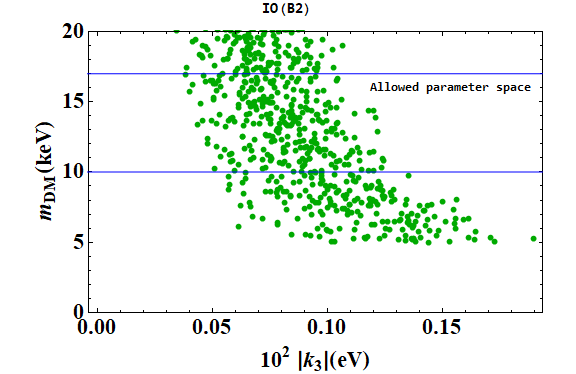}
		\includegraphics[width=0.30\textwidth]{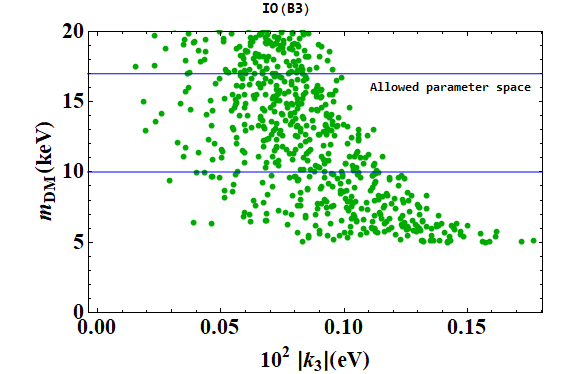}
	\end{center}
	\begin{center}
		\caption{Dark matter mass as a function of model parameter k3 for NO as well as IO}
		\label{fig7}
	\end{center}
\end{figure}
\begin{figure}[H]
	\begin{center}
		\includegraphics[width=0.30\textwidth]{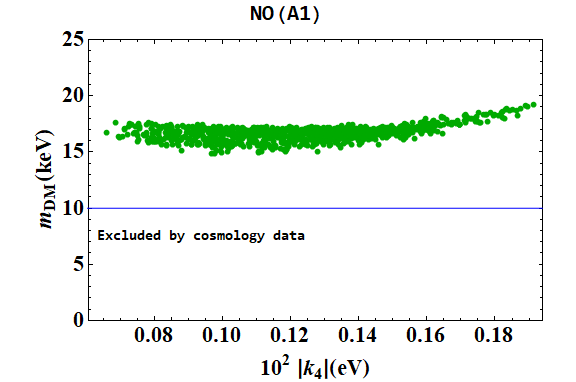}
		\includegraphics[width=0.30\textwidth]{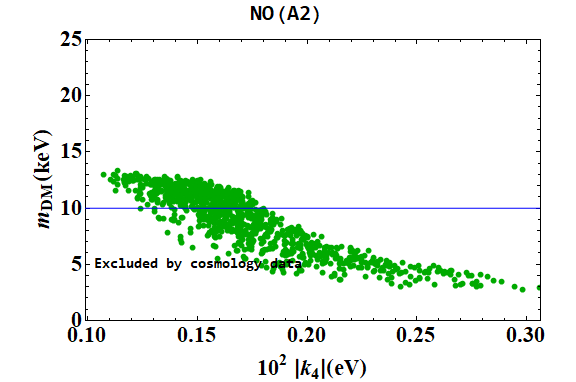}
		\includegraphics[width=0.30\textwidth]{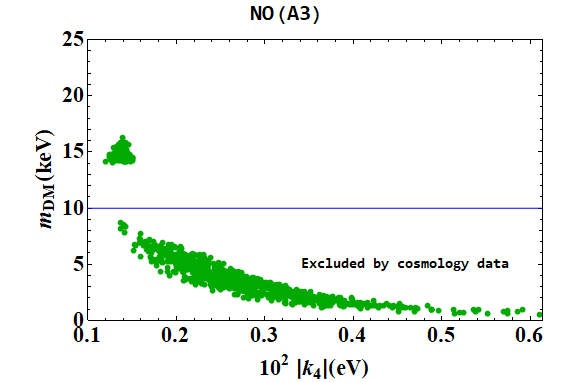}\\
		\includegraphics[width=0.30\textwidth]{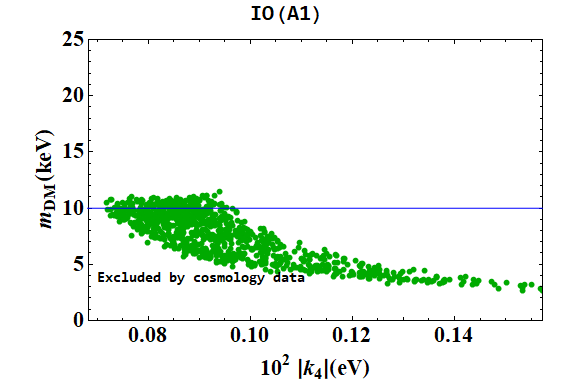}
		\includegraphics[width=0.30\textwidth]{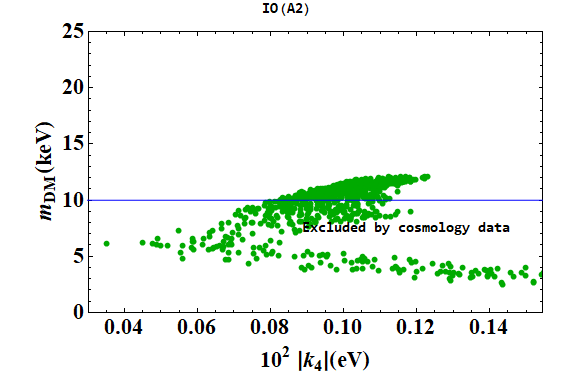}
		\includegraphics[width=0.30\textwidth]{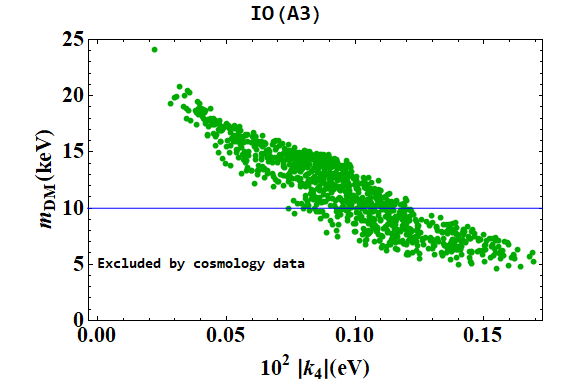}\\
		\includegraphics[width=0.30\textwidth]{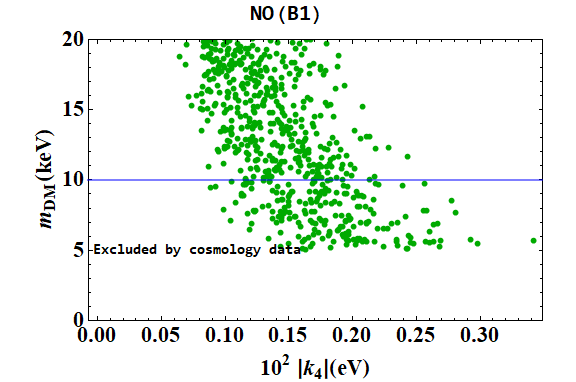}
		\includegraphics[width=0.30\textwidth]{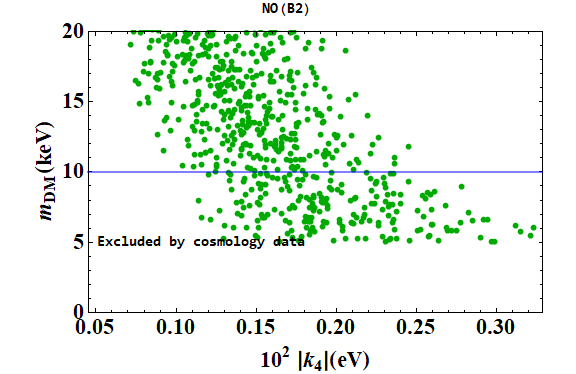}
		\includegraphics[width=0.30\textwidth]{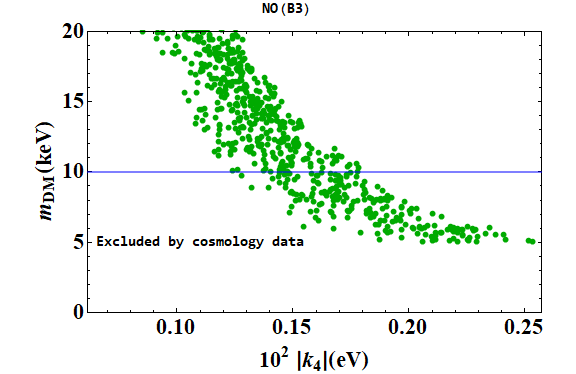}\\
		\includegraphics[width=0.30\textwidth]{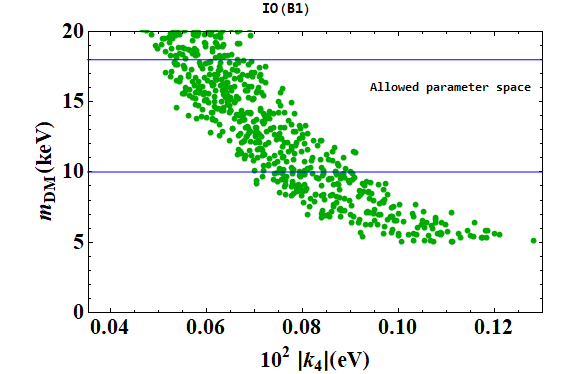}
		\includegraphics[width=0.30\textwidth]{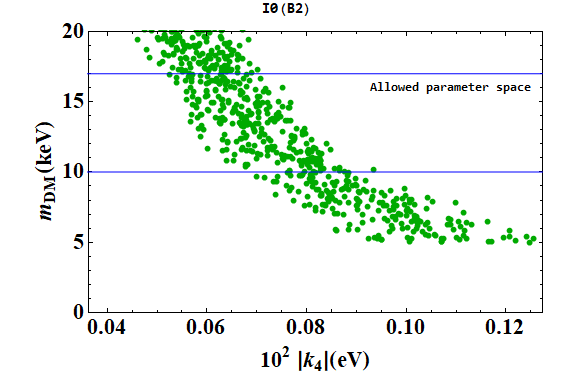}
		\includegraphics[width=0.30\textwidth]{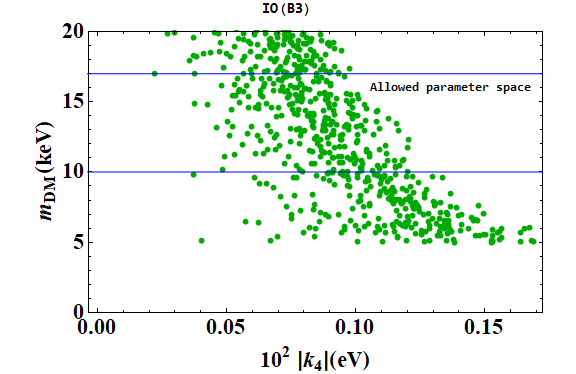}
	\end{center}
	\begin{center}
		\caption{Dark matter mass as a function of model parameter k4 for NO as well as IO}
		\label{fig8}
	\end{center}
\end{figure}
From fig \ref{fig1}, it is evident that texture A1 yields very small DM-active mixing. X-ray data do not have any bound on mass below 20 keV in this range of mixing. Again, the mass range predicted by A1 is above the lower mass bound coming from Ly- $\alpha$ data. Thus A1 leads to such mass and mixings which are all allowed by cosmology data. Similarly, a very small mass-mixing parameter space spanning from 10 to 14 keV allowed in texture A2. Also, X-ray data do not possess any bounds on mass because of the small mixing. In texture A3, a quite small parameter space is available considering the cosmological bounds. In the textures, B1, B2 and B3, the allowed mass-mixing parameter space lies within $10-20$ keV. Previously in our work \cite{Gautam:2019pce}, the texture with non vanishing entries leads to the bound on sterile neutrino dark matter mass as $10-17$ keV. But with texture zeros in $M_{D}$, the mass range above 17 keV is also allowed as shown in the case of B1,B2 and B3. Again, we get the bound on the upper limit on the mass of sterile neutrino as 14 keV (in A1 and A2) which is lower than the previously obtained limit. It is observed in fig \ref{fig1a} that the bounds on mass of the sterile neutrino dark matter obtained from texture A1 and A2 considering IO are 12 keV and 13 keV respectively which are lower than the previously obtained bound in \cite{Gautam:2019pce}. A3 yields the upper bound on dark matter mass as 18 keV. However, in the case of textures, B1,B2, and B3, the upper limit on the mass of the dark matter is 17 keV. Thus, we can see from our analysis that the allowed range of the mass of dark matter is highly dependent on the different two zero textures of $M_{D}$. 

The decay rate of the proposed DM candidate in the six categories is shown in fig \ref{fig2} and fig \ref{fig2a} for NO and IO respectively. It is observed that the decay rate is very small in both cases ensuring the stability of the proposed dark matter in all the categories.

Fig \ref{fig3} and fig \ref{fig3a}represent the relic abundance of the dark matter in all the categories for NO and IO respectively. It has been already mentioned in \cite{Abada:2014zra}, the sterile neutrino has a partial contribution to total dark matter relic abundance. In our analysis, it is observed that the relic abundance is dependent on the structures of $M_{D}$. Different two zero textures of $M_{D}$ lead to distinguished relic abundance. The textures A1, A3, and B2 in NO yield relic abundance up to $83\%$ of the total DM abundance for the allowed mass range of the dark matter in these cases. On the other hand, the relic abundance obtained from the textures A2 and B1 are quite low and can account for only $15\%$ of the total DM abundance for NO. Again, in the case of IO, texture A1 contributes to $90\%$ of the total DM abundance for the allowed mass range as can be seen from fig \ref{fig3a}. This result improves the contribution of relic abundance obtained previously without considering any zero textures in \cite{Gautam:2019pce}. However, the textures A2, A3, B2, and B3 produce an overabundance of the dark matter along with a small amount of proper relic in the case of IO.   

We have shown the effective neutrino mass in $0\nu\beta\beta$ with contributions of sterile neutrino in the framework of ISS (2,3) in fig \ref{fig4}. In all the classes, we have compared the results with the experimental limit provided by different experiments. From fig \ref{fig4}, we can conclude that both NO and IO satisfy the experimental bounds for all the categories despite the presence of heavy neutrinos in the model.

Fig \ref{fig5} to fig \ref{fig8} represent the dark matter mass as a function of the four model parameters. We have implemented the current cosmological bounds on the mass of dark matter to study the range of model parameters k1,k2,k3, and k4 satisfying the limits in all the textures. These parameters correspond to the non-zero entries in $M_{\nu}$. The range of allowed parameters are summarised in table \ref{tab5}.
\begin{table}[H]
	\centering
	\resizebox{\columnwidth}{!}{%
	\begin{tabular}{|c|c|c|c|c|}
	\hline 
		Texture  & k1 (NO/IO) eV & k2 (NO/IO) eV & k3 (NO/IO) eV & k4 (NO/IO) eV \\ 
		\hline 
		A1 & $(0.02-0.2)/(0.15-0.17)$ & $(0.02-0.2)/(0.09-0.14)$  &$(0.07-0.25)/(0.05-0.1)$ &$(0.08-0.2)/(0.07-0.1)$ \\
		\hline
		A2 & $(0.01-0.03)/(0.18-0.25)$ & $(0.03-0.07)/(0.04-0.10)$  &$(0.1-0.2)/(0.06-0.1)$  &$(0.11-0.18)/(0.09-0.13)$ \\
		\hline 
		A3 & $(0.03-0.06)/(0.1-0.2)$ & $(0.01-0.1)/(0.19-0.24)$  &$(0.1-0.2)/(0.05-0.14)$  &$(0.04-0.06)/(0.15-0.25)$ \\
		\hline 
		B1  &  $(0.12-0.17)/(0.14-0.2)$  & $(0.1-0.16)/(0.04-0.12)$  &$(0.01-0.2)/(0.05-0.14)$  &$(0.12-0.17)/(0.14-0.20)$ \\
		\hline
		B2  & $(0.13-0.19)/(0.12-0.17)$  & $(0.04-0.12)/(0.04-0.12)$  &$(0.07-0.09)/(0.05-0.13)$  &$(0.12-0.18)/(0.14-0.2)$ \\
		\hline
		B3  & $(0.03-0.06)/(0.10-0.15)$  & $(0.01-0.06)/(0.1-0.3)$  &$(0.1-0.17)/(0.05-0.13)$  &$(0.03-0.07)/(0.10-0.17)$ \\
		\hline
	\end{tabular} %
}
	\caption{The range of model parameters satisfying the dark matter mass within the cosmological bound.} \label{tab5}
\end{table} 
\begin{table}[H]
	\centering
	\begin{tabular}{|c|c|c|c|}
		\hline 
	    Texture  & NDBD & Relic abundance & Decay rate \\ 
		\hline 
		A1 (NO/IO) & $\checkmark (\checkmark)$ &$\checkmark (\checkmark)$ &$\checkmark (\checkmark)$\\
		\hline
		A2 (NO/IO) & $\checkmark (\checkmark)$ & $\times(\times)$ &$\checkmark (\checkmark)$\\
		\hline 
		A3 (NO/IO) & $\checkmark (\checkmark)$ &  $\checkmark(\times)$ & $\checkmark (\checkmark)$\\
		\hline 
		B1 (NO/IO) & $\checkmark (\checkmark)$ &  $\times(\times)$  &$\checkmark (\checkmark)$\\
		\hline
		B2 (NO/IO) & $\checkmark (\checkmark)$ &$\checkmark (\times)$&$\checkmark (\checkmark)$\\
		\hline
		B3 (NO/IO) & $\checkmark (\checkmark)$ & $\checkmark (\times)$  &$\checkmark (\checkmark)$\\
		\hline
	\end{tabular} 
	\caption{Summary of allowed and disallowed textures.} \label{tab6}
\end{table} 
In the previous work on maximal zeros in inverse seesaw framework \cite{Adhikary:2013mfa,Sinha:2015ooa}, the authors have considered all the three neutrino masses to be non zero. The lightest neutrino mass obtained in \cite{Adhikary:2013mfa} is (0.003-0.086) depending on the textures. In our work with minimal inverse seesaw, the lightest neutrino mass is zero and it is the special feature of the model which further constrains the other two masses of the active neutrinos.  
The maximal zeros in inverse seesaw framework predicts that the number of maximal zeros allowed in Dirac mass matrix ($M_{D}$) is four to give neutrino phenomenology and successful lepton asymmetry \cite{Adhikary:2013mfa}. It can be concluded from our work that the maximum number of zeros allowed in $M_{D}$ in minimal inverse seesaw is two. Within inverse seesaw, it is quite interesting that some of the textures can lead to NO while others can lead to IO in \cite{Adhikary:2013mfa}. Again, texture zeros in inverse seesaw may lead to normal ordering of the neutrino mass as mentioned in \cite{Sinha:2015ooa}. In the paper \cite{Ghosal:2015lwa}, the ordering obtained in all the cases are inverted which is an important result of inverse seesaw with maximal zeros. However, ISS(2,3) with maximal zeros leads to both normal and inverted ordering. It does not completely rule out any ordering of light neutrino masses from the neutrino phenomenology. However, three textures with IO are not allowed from dark matter phenomenology. The significance of ISS(2,3) with maximal zeros is that it can further put constraints on sterile neutrino dark matter mass which will hopefully be detected in the obtained mass range in near future.
\section{\label{sec:level7}Searches for sterile neutrino dark matter in present and future experiments}
Owing to the importance of sterile neutrino in particle physics as well as cosmology, different techniques have been adopted to search for the sterile neutrinos in the keV scale. These techniques can be categorised as direct detection by using large scale detectors and indirect detection by observing signatures of dark matter in some laboratory experiments \cite{Divari:2017coh}. There may be significant impacts of sterile neutrino in keV scale on $0\nu\beta\beta$ which improves the rate of such process \cite{Abada:2018qok,Rodejohann:2014eka}. In our study, all the texture zeros of mass matrices comply with the KamLAND-Zen data and we do not observe significant contributions to $0\nu\beta\beta$ because of the mass range. However, the impact of
sterile neutrino dark matter on most other observables can provide a way to indirectly search for such candidates in $0\nu\beta\beta$ experiments in the future. This search can be used to indirectly test the different zero textures proposed in our work. 

As proposed in \cite{Boyarsky:2018tvu,Long:2014zva,Shaposhnikov:2007cc}, it may be possible to detect keV neutrino through $\nu$-capture on $\beta$ decaying nuclei. One of these involves counting the number of $^{163}$Ho atoms in rare-earth ores from the $\nu$-capture on $^{163}$Dy \cite{Boyarsky:2018tvu,Lasserre:2016eot}. The results shown in \cite{Boyarsky:2018tvu}, indicates that a minimum value of the DM-active mixing $sin^{2}\theta_{e}\sim 10^{-6}$ is attainable with a kg-scale target mass. But this value is not sufficient to comply with cosmological and astrophysical data. In \cite{Lasserre:2016eot}, authors have mentioned that there is possibility to explore mixing angles as low as $sin^{2}\theta_{e}\sim 10^{-7}$ with a real-time experiment. As mentioned in \cite{Boyarsky:2018tvu}, lower mixing angles $sin^{2}\theta_{e}\sim 10^{-9}$ could in principle be explored with 100 tons of $^{163}$Dy with detector backgrounds less than $10^{6}$ counts/kg/day/keV. We can also expect new proposals for laboratory experiments that improve the previously obtained limits and may reach a region of astrophysical interest shortly. 

Another direct detection technique involves in search for the scattering of sterile neutrinos having some mixing with active neutrinos in large detectors. XENON100, XENON1T, and DARWIN are some of these experiments  \cite{Aprile:2012nq}. XENON1T experiment reaches a mass range between 10 and
40 keV of sterile neutrinos \cite{Aprile:2012nq}. Most of the textures discussed above are within the reach of experiments (as the dark matter mass predicted is within $10-40$ keV), provided these can lead to mixing angle $sin^{2}\theta_{e}\sim 10^{-6}$ explored by DARWIN experiment.

It is also possible to detect sterile neutrino by searching for its impact on $\beta$ decay spectra via a characteristic distortion in it. KATRIN \cite{Angrik:2005ep}, Electron Capture on Holium (ECHo) \cite{Gastaldo:2013wha}, TRISTAN \cite{Mertens:2014nha}, Ptolemy \cite{Betts:2013uya} are some of the present and planned experiments that aim at searching for sterile neutrinos in keV scale in $\beta$ decay spectrum.

Indirect detection of sterile neutrinos involves searches for X-ray emission and structure formation. We have implemented the already existing bounds in all the textures mentioned above. However, we would expect significant improvement of the data by future optical telescope and also from different X-ray experiments like ATHENA or XARM. Considering all these possibilities of detecting sterile neutrino in keV scale, we have chosen to study the impacts of texture zeros on the phenomenology of this particular DM candidate. We expect that the results from the experiments in near future would also increase the predictability of the textures. 
 
\section{\label{sec:level8}Conclusion}
In this work, we have explored the scenario with only two right-handed neutrinos and texture zeros of the mass matrices considering the diagonal charged lepton mass matrix. We have performed a thorough analysis of texture zeros of Dirac mass matrix $M_{D}$ as it plays an important role in generating light neutrino mass matrix in the framework of inverse seesaw ISS(2,3). In this framework, $M_{D}$ is a non squared $2\times3$ matrix and maximal zeros are imposed in the other two matrices $M_{N}$ and $\mu$. Considering the allowed texture zeros in the light neutrino mass matrix, we find out all texture zero possibilities in Dirac mass matrix. One of the main points of this work is that with the maximal zero textures of $M_{N}$ and $\mu$, the maximum allowed number of zeros in $M_{D}$ is two to give correct neutrino phenomenology. With the six allowed classes of two zero textures of $M_{D}$, we have studied dark matter phenomenology and constrain the texture zero
mass matrices from the relevant cosmological bounds. We have also evaluated the contributions to effective Majorana neutrino mass ($m_{ee}$) characterising 0$\nu\beta\beta$ in all the textures. Table \ref{tab6} represents the summary of our results. It has been observed that all the textures are allowed from the relevant experimental bounds on $m_{ee}$. However, some two zero textures are disallowed from the cosmology data. It is evident that the textures A2 (NO/IO), A3(IO), B1 (NO/IO), B2(IO), and B3(IO) lead to unacceptable values of relic abundance of the dark matter. Thus there are only four Dirac matrices in NO yield predictions allowed by the cosmology data. Again another important point of the present work is that the bounds on sterile neutrino dark matter is dependent on the texture zeros of $M_{D}$ and yield upper bounds on sterile neutrino dark matter mass different from the one with non vanishing textures of $M_{D}$ as in earlier work \cite{Gautam:2019pce}. The upper limit predicted by zero textures is $20$ keV in NO and that for the IO is $18$ keV. Another important finding is one of the classes of texture zeros predicts relic abundance up to $90\%$ of the total abundance which improves the previously obtained result without considering texture zeros. In summary, though there are six allowed classes of texture zero models in the framework of ISS$(2,3)$, yet some of these textures are astonishingly weak in their predictions. However, other textures improve the contribution of sterile neutrino to the total dark matter abundance and also put stronger bounds on sterile neutrino dark matter mass. We leave the detailed study of the flavor symmetric origin of such textures and related phenomenology to future works.
\section*{Acknowledgements}
NG acknowledges Department of Science and Technology (DST),India(grant DST/INSPIRE Fellowship/2016/IF160994) for the financial assistantship. The work of MKD is supported by the Department of Science and Technology, Government of India under the project no. $EMR/2017/001436$.

\bibliographystyle{paper}
\bibliography{texture}
\end{document}